\documentclass[conference]{IEEEtran}
\usepackage[utf8]{inputenc}
\usepackage[T1]{fontenc}
\usepackage{graphicx}      
\usepackage{amsmath,amssymb}
\usepackage{booktabs}      
\usepackage{array}
\usepackage{threeparttable}
\usepackage{float}         
\usepackage{listings}      
\usepackage{xcolor}        
\usepackage{url}           
\usepackage{cite}
\usepackage{enumitem}
\usepackage{multirow}
\usepackage{placeins}

\graphicspath{{images/}}

\usepackage[caption=false,font=footnotesize]{subfig}

\usepackage[hidelinks]{hyperref}

\IEEEoverridecommandlockouts
\title{A Critical Analysis of the Medibank Health Data Breach and Differential Privacy Solutions} 
\author{
	\IEEEauthorblockN{Zhuohan Cui$^{*}$}
	\IEEEauthorblockA{The University of Sydney}
	\and
	\IEEEauthorblockN{Qianqian Lang}
	\IEEEauthorblockA{Dalian Neusoft University of Information}
	\and
	\IEEEauthorblockN{Zikun Song}
	\IEEEauthorblockA{The University of Sydney}
	\thanks{$^{*}$Corresponding author: Zhuohan Cui (email: zcui0321@uni.sydney.edu.au).}
}       

\begin{document}
\maketitle
\begin{abstract}
	This paper critically examines the 2022 Medibank health insurance data breach, which exposed sensitive medical records of 9.7 million individuals due to unencrypted storage, centralized access, and the absence of privacy-preserving analytics. To address these vulnerabilities, we propose an entropy-aware differential privacy (DP) framework that integrates Laplace and Gaussian mechanisms with adaptive budget allocation. The design incorporates Transport Layer Security (TLS)-encrypted database access, field-level mechanism selection, and smooth sensitivity models to mitigate re-identification risks. Experimental validation was conducted using synthetic Medibank datasets (N = 131,000) with entropy-calibrated DP mechanisms, where high-entropy attributes received stronger noise injection. Results demonstrate a 90.3\% reduction in re-identification probability while maintaining analytical utility loss below 24\%. The framework further aligns with GDPR Article~32 and Australian Privacy Principle~11.1, ensuring regulatory compliance. By combining rigorous privacy guarantees with practical usability, this work contributes a scalable and technically feasible solution for healthcare data protection, offering a pathway toward resilient, trustworthy, and regulation-ready medical analytics.
\end{abstract}


\section{Introduction}
\label{sec:Introduction}

\subsection{Understanding Data Privacy}
\label{subsec:}
Data privacy refers to the right of individuals to control their data. Most companies regularly collect, store, and use user data in their daily operations. This data may include PII (personally identifiable information), such as name, phone number, and email address. Data privacy requires companies to obtain user consent before using their data, implement security measures to protect stored data from disclosure, and empower users to manage their own collected data \cite{ibm2025}. While privacy principles mandate encryption and user control, the Medibank breach illustrates how failure to operationalize these principles leads to catastrophic exposure of sensitive medical records.

\subsection{The Escalating Scale of Data Breaches}
In recent years, the number and scale of global data breaches have continued to rise. Research indicates that cybercrime losses are expected to exceed \$12.5 billion in 2023, nearly double the amount reported in 2021. The average global cost of a data breach in 2023 has risen to \$4.45 million, and the cost of cybercrime is projected to reach \$10.5 trillion by 2025. Furthermore, the costs of data breaches extend far beyond monetary value, posing even more serious challenges, such as impacting user psychological well-being, corporate reputation, and social trust \cite{gracy2025}. Medibank loses over \$125 million in data breach\cite{medibank2024breach}. Unlike financial breaches where losses are primarily monetary, the Medibank incident demonstrates how healthcare breaches amplify psychological harm and long-term identity risks, underscoring the inadequacy of generic breach statistics in capturing sector-specific vulnerabilities.

\subsection{Literature Review}
The Medibank breach exposed systemic weaknesses in healthcare data protection. To critically evaluate potential solutions, this section reviews three major categories of privacy-preserving technologies—Differential Privacy (DP), Homomorphic Encryption (HE), and Federated Learning (FL)—and assesses their relevance to the vulnerabilities observed in Medibank’s architecture.

\subsubsection{Applications of DP in Medical Data}
DP has become a core research direction in medical data protection. It adds calibrated noise to query outputs or gradients to reduce the risk of membership inference.

The following are representative applications of DP in medical data analysis:

\begin{itemize}
	\item \textit{A Survey on DP for Medical Data Analysis:} Prior work demonstrates the application scenarios and challenges of DP in medical data, genomics, and wearable device data, and points out the accuracy loss and privacy budget allocation difficulties faced by DP when processing high-dimensional medical data.\cite{wang2023privacy}.
	
	\item \textit{DP in Health Research:} This review provides a scoping review of the understanding, adoption, and practical application scenarios of DP. It analyzes the potential of DP in clinical data sharing, public health research, and remote health monitoring, and summarizes practical deployment experiences of DP in health research. The study finds that there remain significant differences among medical researchers in understanding and tuning the $\varepsilon$ parameter\cite{mehta2021balancing}. Medibank’s analytics lacked DP, exposing raw statistics; prior work shows DP could have mitigated re-identification.
	
	\item \textit{DP in Medical Imaging Applications:} This book focuses on the specific application of DP in medical imaging, discussing technical solutions for achieving privacy protection through noise injection during the training and inference phases of deep learning models. Research indicates that the high-dimensional nature of image data complicates the balancing of noise mechanisms, necessitating a dynamic trade-off between model accuracy and privacy. Incorporating a layer-wise noise injection mechanism during image model training can enhance privacy protection without significantly compromising diagnostic accuracy.\cite{barthe2024dp}.
	
	\item \textit{Privacy-Preserving ML for electronic health records (EHRs) using FL and DP:} This paper proposes a framework for protecting EHRs that combines FL with DP. By training models locally at each institution and incorporating DP noise, it enables secure data sharing in cross-institutional collaboration. Experiments demonstrate that this scheme effectively mitigates the risk of member inference attacks while maintaining high model performance, providing a practical and feasible privacy-preserving paradigm for distributed learning of medical data.\cite{kairouz2024landscape}. Medibank’s centralized architecture created a single point of failure; federated approaches could have avoided this concentration of risk.

	\item \textit{Leveraging FL for Privacy-Preserving Analysis of Multi-institution EHRs:} This book further expands on the application of DP in collaborative analysis of multi-institutional EHRs, proposing a mechanism for dynamically allocating privacy budgets within a FL pipeline to account for differences in data size and model contributions across institutions. Emphasizing DP not only enhances trust in cross-organizational data sharing, but also provides technical support for compliance with privacy regulations such as GDPR and HIPAA.\cite{li2024advances}.
\end{itemize}

Medibank’s analytics lacked DP, exposing raw statistics; prior work shows DP could have mitigated re-identification.

\subsubsection{The Latest Developments in Data Loss Prevention}
\begin{itemize}
	
	\item {The technological evolution for medical data leakage prevention is shifting from ``perimeter protection + traditional encryption'' to ``multi-layered, synchronized defense + end-to-end privacy protection'':}
	In recent years, with the surge in medical data volumes and the frequent occurrence of cyberattacks and data leaks, single protection mechanisms (such as firewalls or traditional encryption) have become inadequate to meet the security requirements of data accessed by multiple devices, in multiple scenarios, and by multiple roles. Research indicates that multi-layered defense—that is, the simultaneous deployment of security mechanisms at multiple levels, including physical, network, application, data access, and output auditing—is becoming mainstream. Furthermore, privacy-enhancing computing (PEC) technologies (such as DP, HE, secure multi-party computation (SMPC), and trusted execution environments (TEEs)) enable data to maintain a high level of privacy during processing, analysis, and sharing.
	For example, one study indicates that key privacy-enhancing technologies (PETs) in collaborative medical analytics include FL, DP, HE, SMPC, and the convergence of these technologies is accelerating~\cite{orca2025}.
	
	\item {Google's medical AI platform uses Federated Analytics + DP Aggregation.}  
	By eliminating centralized data exchange and adopting a ``model/statistical output sharing + DP protection'' mechanism, this architecture reduces data leakage risks and supports regulatory compliance \cite{googlefederated}.  
	Medibank, by contrast, relied on centralized data storage and analytics without DP, leaving raw statistics exposed and creating a single point of failure. This comparison highlights how industry leaders are already deploying federated and DP-based architectures to mitigate risks that Medibank failed to address.

	\item {IBM Homomorphic Encryption Framework:}
	HE is a technique that allows calculations (such as addition and multiplication) to be performed directly on encrypted data without first decrypting it, thereby ensuring the confidentiality of the data throughout the entire processing process. The medical field is particularly suitable for this type of technology because the data is highly sensitive and there is a strong demand for analysis. A recent review pointed out that the application of HE in medical data processing (such as EHR, genetic data, and medical imaging) is accelerating. Related research covers types such as partially homomorphic encryption, somewhat homomorphic encryption, and fully homomorphic encryption, and analyzes attack surfaces and defense mechanisms~\cite{springerhe}.
	For example, IBM and its research team launched a HE toolkit specifically for FL scenarios~\cite{ibmfl}. One of the goals is to enable multiple medical institutions to conduct joint modeling or analysis while keeping the data encrypted. In this way, data can still participate in large-scale collaborative analysis without leaving the local area, thereby enhancing the value of data while protecting privacy. Medibank processed sensitive data in plaintext; HE frameworks demonstrate how encrypted computation could have prevented exposure.
	
	\item {Encrypted Computation allows data to remain confidential even during processing:}
	In traditional approaches, data is encrypted during storage or transmission but typically needs to be decrypted for computation, which exposes it to potential leakage. Privacy‑enhancing computing technologies such as HE \cite{gentry2009fully}, TEEs \cite{sabt2015trusted}, and SMPC \cite{yao1982protocols} enable data to remain encrypted or protected even during the processing stage. These methods significantly reduce the risk of data leakage during computation and have become foundational techniques for secure analytics in healthcare and other sensitive domains. Medibank’s reliance on plaintext processing without encrypted computation illustrates how the absence of these privacy-enhancing technologies directly contributed to its systemic vulnerability.
\end{itemize}

\subsubsection{Comparative Study of Privacy-Preserving Techniques}
In modern medical data analysis, privacy protection has become a core requirement for algorithm design and system deployment. Prior research has emphasized that data confidentiality must be preserved throughout the computational process and that raw data should not be directly accessible to any participant \cite{dwork2006calibrating,goldwasser1984probabilistic,yao1982protocols}. Current research focuses on three major categories of privacy-enhancing technologies: DP, HE, and SMPC. Table~\ref{tab:privacy-techniques} contrasts DP, HE, and SMPC. Medibank’s failure to deploy any of these mechanisms underscores the systemic weakness of its privacy architecture.

\begin{table}[htbp]
	\caption{Comparison of Core Privacy-Enhancing Techniques}
	\label{tab:privacy-techniques}
	\centering
	\begin{threeparttable}
		\renewcommand{\arraystretch}{1.2}
		\setlength{\tabcolsep}{3pt}
		\footnotesize
		\begin{tabular}{|>{\centering\arraybackslash}m{1.7cm}|
				>{\centering\arraybackslash}m{2.1cm}|
				>{\centering\arraybackslash}m{2.1cm}|
				>{\centering\arraybackslash}m{2.1cm}|}
			\hline
			\textbf{Technique} & 
			\textbf{Basic Principle} & 
			\textbf{Main Advantages} & 
			\textbf{Main Limitations} \\ \hline
			
			Differential Privacy (DP) & 
			Adds controlled random noise to statistical results or model gradients so that the participation or absence of any single record does not significantly affect the output. &
			Low computational overhead; simple to implement; suitable for large-scale statistical analysis and model training with privacy constraints. &
			Excessive noise may reduce accuracy; difficult to balance privacy budget ($\varepsilon$) and model performance. \\ \hline
			
			Homomorphic Encryption (HE) &
			Enables mathematical operations (addition, multiplication) directly on encrypted data, producing the same results as computations on plaintext after decryption. &
			Maintains data confidentiality during storage and computation; minimal accuracy loss; ideal for secure cloud training and inference. &
			High computational complexity and latency; heavy resource requirements for edge devices. \\ \hline
			
			Secure Multi-Party Computation (SMPC) &
			Splits data into random shares distributed across multiple parties, allowing joint computation through secure protocols without revealing individual inputs. &
			Allows collaborative modeling without centralizing data; inherently supports multi-institutional cooperation. &
			Requires multiple communication rounds; high synchronization and latency overhead; protocol implementation can be complex. \\ \hline
		\end{tabular}
	\end{threeparttable}
\end{table}
In summary, while DP, HE, and FL+DP frameworks have been extensively studied and deployed in healthcare contexts, Medibank’s breach reveals a critical gap between research and practice. The absence of these mechanisms directly enabled re-identification, plaintext exposure, and systemic failure, highlighting the urgent need for entropy-aware DP integration.

\subsection{Future Implications}
\subsubsection{Trends in AI-driven Privacy Attacks}
With the rapid adoption of generative AI, the forms and sophistication of privacy attacks are rapidly evolving. Traditional attacks primarily focused on direct data leakage or parameter extraction, but today's attackers are leveraging the generative and inference capabilities of large models to more covertly mine information stored within them. Medibank’s lack of DP makes it particularly vulnerable to membership inference attacks, which are increasingly automated by generative AI.

\begin{itemize}
	\item {Membership Inference Attack (MIA) Mechanism:}  
	Deep learning models are prone to memorizing sensitive training data unless privacy-preserving techniques are applied \cite{shokri2017membership}. The core idea of MIA is that an adversary can determine whether a sample belongs to the model’s training set by analyzing confidence scores, gradients, or probability distributions of the model’s outputs. Because deep models often exhibit overfitting or “memorization,” attackers can infer whether the model has seen the real data of a patient, user, or customer using only a few queries. In recent years, MIA has expanded beyond classification tasks to include generative models and time-series forecasting \cite{hu2021membership}, posing particular risks to healthcare and financial sectors where sensitive data is widely used.
	
	\item {Large Model and “Shadow Data” Attacks:}  
	In the future, attackers may leverage large language models (LLMs) or diffusion models to generate “shadow data.” These synthetic samples can then be used to test the target model's response patterns, thereby inferring the distribution of the original training set or the content of specific samples.  
	The danger of this approach lies in the fact that attackers do not need direct access to the training data. Instead, they interact with the model in a “black box” manner, leveraging the LLM's strong generative and pattern matching capabilities to reconstruct or approximate the original data features.  
	For example, in healthcare scenarios, attackers could use constructive prompt injection or continuous questioning to cause the model to leak protected case descriptions, pathological features, or the source of the corpus.
\end{itemize}

\subsubsection{Quantum Threats to Cryptography}
Classic HE algorithms, including the Goldwasser--Micali (GM) scheme \cite{goldwasser1984probabilistic} and the Brakerski--Fan--Vercauteren (BFV) scheme \cite{brakerski2012leveled}, rely on traditional mathematical problems such as integer factorization and discrete logarithms. Under the classical computing model, these problems are considered computationally hard, thereby guaranteeing data confidentiality during computation and storage. However, the emergence of quantum computing is changing this assumption, as algorithms such as Shor's algorithm \cite{shor1994algorithms} can efficiently solve these problems, posing systemic risks to long-term cryptographic security. Medibank's reliance on conventional encryption---and the absence of post-quantum cryptography (PQC)---raises long-term risks of ``harvest now, decrypt later'' attacks.

\begin{itemize}
	\item {The disruptive impact of quantum computing on traditional encryption:}  
	Quantum computers, based on the principles of superposition and entanglement, can execute specific algorithms in an exponentially parallel state space. The most representative quantum algorithm is Shor's algorithm (1994), which can efficiently factor large integers and compute discrete logarithms in polynomial time, posing a direct threat to RSA, Diffie--Hellman, ElGamal, and HE schemes based on these problems. Breakthroughs in quantum computing could potentially crack currently used public-key systems within hours or even minutes, posing systemic risks to long-term data storage systems in healthcare, finance, and government sectors. Medibank’s reliance on conventional encryption without quantum-resistant safeguards illustrates how such breakthroughs could transform today’s plaintext vulnerabilities into tomorrow’s systemic failures.

	\item {Unique Vulnerabilities in Medical Data Security:}  
	The medical industry retains data for long periods of time (often over 30 years) and is highly sensitive to privacy (involving genetic and medical history information). This means that even if attackers cannot decrypt the data currently, they can employ a “harvest now, decrypt later” strategy: stealing encrypted backups in advance and decrypting them later when quantum computing matures. This delayed attack model poses a long-term vulnerability to EHRs, genetic data warehouses, and photographic archiving systems (PACS). If this data is cracked by quantum computing, it will not only expose personal privacy but also potentially impact insurance assessments and the reproducibility of medical research results, and could even be used for model poisoning or identity forgery attacks.
	
	\item {PQC and Quantum-Safe Homomorphic Encryption (Quantum-Safe HE):}  
	To address the quantum threat, international standards bodies such as NIST have launched a PQC algorithm standardization program, focusing on cryptographic systems based on quantum-resistant assumptions such as lattice-based, code-based, multivariate polynomial, and hash-based cryptography.  
	In HE, researchers are exploring parameter enhancement schemes for LWE (Learning With Errors) and RLWE (Ring-Learning With Errors), as well as hybrid HE + PQC protocols, to achieve quantum security while maintaining computational efficiency.
\end{itemize}

\subsubsection{Privacy Challenges in Edge and Federated Environments}
With the increasing prevalence of wearable devices, home medical sensors, and remote monitoring terminals, healthcare systems are evolving from centralized architectures toward edge computing and FL models. These devices can collect high-frequency health data such as heart rate, blood sugar, body temperature, and movement trajectory in real time, supporting personalized diagnosis and treatment, as well as intelligent intervention. However, these edge nodes typically have limited computing and storage resources, operate in complex environments, and experience unstable network conditions, making traditional encryption and privacy protection mechanisms challenging to deploy.

FL has been shown to enhance privacy guarantees by allowing each node to train models locally without centralizing the original data \cite{mcmahan2017communication,yang2019federated}. This mechanism reduces the risk of data exposure while enabling collaborative model building across distributed environments. Medibank’s centralized model ignored federated alternatives, leaving it exposed to systemic breach rather than localized containment.

However, prior studies also highlight two key bottlenecks:

\begin{itemize}
	\item High communication overhead — frequent model parameter synchronization and gradient uploads consume significant bandwidth and energy;
	\item Model synchronization challenges — node heterogeneity and latency variations make it difficult for the global model to converge stably.
\end{itemize}

These issues make edge privacy protection no longer simply a design challenge at the algorithm level; it also involves optimizing system architecture and resource scheduling. Lightweight differential privacy (LDP) and federated homomorphic encryption (FHE) offer inherent advantages in privacy protection.

\begin{itemize}
	\item {Lightweight Differential Privacy (LDP):}  
	Traditional DP algorithms inject noise during each model update or aggregation phase. While this can prevent individual samples from being re-identified, it consumes significant computational and bandwidth resources, making it unsuitable for resource-constrained end-user environments.  
	LDP utilizes strategies such as local perturbation, gradient compression, and adaptive noise injection to allocate privacy budgets from the global level down to the node level, significantly reducing computational complexity and energy consumption. For example, research has shown that LDP, which combines pruning and sparsification, can reduce communication traffic by over 50\% while maintaining model performance.  
	In medical applications, this approach allows wearable devices to perform privacy-preserving processing on ECG or sleep data locally and then upload denoised features, thus balancing privacy and real-time performance.
	
	\item {Federated Homomorphic Encryption (FHE):}  
	The core idea of federated FHE is:  
	\begin{itemize}
		\item Model updates are encrypted using a lightweight FHE library on each device;
		\item A federated aggregation server merges the models in the ciphertext domain;
		\item The results are then broadcast back to each node for decryption and the next round of training.
	\end{itemize}
	To reduce computational costs, recent research has proposed optimizations such as ``partially homomorphic aggregation'' and ``batch encoding,'' which have reduced FHE inference time on mobile devices by one-third. Medibank, however, processed sensitive medical data in plaintext within a centralized architecture; federated FHE could have prevented systemic exposure by ensuring encrypted computation and distributed aggregation.
	
\end{itemize}

\subsection{Updates to Laws and Regulations}
Table \ref{tab:lawupdates} indicates that data privacy-related laws and regulations are also being continuously updated. These updates highlight a global trend toward strengthening individual rights, mandating encryption and consent management, and imposing stricter accountability on organizations handling sensitive data. For example, the 2024 Privacy and Other Legislation Amendment in Australia introduces direct legal remedies for individuals affected by serious privacy violations, while ISO/IEC TS 27560:2023 establishes standardized consent receipts to ensure traceability and verifiability of user permissions.

Taken together, these regulatory developments complement the technical safeguards discussed earlier in Table~\ref{tab:privacy-techniques}. While DP, HE, and SMPC each provide distinct mechanisms for protecting medical data, their adoption must be aligned with evolving legal requirements. In healthcare contexts, entropy-aware DP frameworks are particularly well-suited because they balance analytical utility with strong privacy guarantees, while also satisfying mandates under GDPR Article~32 and Australian Privacy Principle~11.1. By contrast, HE and SMPC offer stronger confidentiality but face scalability and latency challenges in real-world medical systems. 

In summary, the convergence of updated privacy regulations (Table~\ref{tab:lawupdates}) and advanced privacy-enhancing technologies (Table~\ref{tab:privacy-techniques}) underscores that effective healthcare data protection requires both technical innovation and regulatory compliance. Medibank’s failure illustrates the risks of neglecting this dual alignment, whereas the proposed entropy-aware DP framework demonstrates a pathway toward resilient, regulation-ready medical analytics.

\begin{table}[htbp]
	\caption{Data Privacy Law Updates}
	\label{tab:lawupdates}
	\centering
	\begin{threeparttable}
		\renewcommand{\arraystretch}{1.1} % 调整行距，美观一点
		\setlength{\tabcolsep}{3pt}       % 调整列间距，避免太宽
		
		\begin{tabular}{|>{\centering\arraybackslash}m{3cm}|
				>{\centering\arraybackslash}m{1.5cm}|
				>{\centering\arraybackslash}m{4cm}|}
			\hline
			\textbf{Regulations} & \textbf{Effective Time} & \textbf{Major Updates} \\ \hline
			
			Privacy and Other Legislation Amendment & 2024 &
			\begin{enumerate}
				\item  Individuals can initiate legal action against organisations or individuals who commit serious privacy violations.
				\item  Sharing another person's information with the intent to harm them is a violation of the law.
				\item The OAIC needs to establish a standard for children's online privacy.
				\item Reasonable measures must be taken to protect personal information \cite{lexology2025}.
			\end{enumerate}
			\\ \hline
			
			ISO/IEC TS 27560:2023 & 2023 &
			Companies must record whether users agree, decline, or withdraw consent. They must also issue users a "Consent Receipt" to ensure traceability and verifiability \cite{pandit2024}. 
			\\ \hline
			
		\end{tabular}
	\end{threeparttable}
\end{table}

\subsection{Exploration of Privacy Anomalies}
In October 2022, Medibank, one of Australia’s largest health insurers, suffered a catastrophic ransomware attack affecting 9.7 million individuals. Sensitive health data, including mental health diagnoses, abortion records, and alcohol treatment, were exfiltrated and later published on the dark web \cite{oaic2024}, \cite{siganto2024}. 

\textbf{Identified Privacy Anomalies:}

\begin{enumerate}[leftmargin=*, itemsep=2pt, topsep=2pt]
	\item \text{Unencrypted Storage of High-Entropy Medical Data:}
	\begin{itemize}[leftmargin=*, itemsep=0pt, topsep=2pt]
		\item Diagnosis codes, treatment types, and medication details were stored
		without adequate encryption.
		\item These fields possess high data entropy, making them uniquely identifying even without names. 
	\end{itemize}
	\item  \text{Lack of Data Segmentation and Access Control:}
	\begin{itemize}[leftmargin=*, itemsep=0pt, topsep=2pt]
		\item All user data was stored in a centralised architecture, violating principles of least privilege and zero trust. 
	\end{itemize}
	\item  \text{Absence of Privacy-Preserving Analytics:}
	\begin{itemize}[leftmargin=*, itemsep=0pt, topsep=2pt]
		\item Medibank’s internal analytics systems lacked DP, exposing raw statistics to potential misuse.
	\end{itemize}
	
\textbf{Potential Consequences for Users:}

	\begin{itemize}[leftmargin=*, itemsep=0pt, topsep=2pt]
		\item Psychological Harm: Public exposure of abortion or mental health records can lead to stigma and trauma. 
		\item Identity Theft: Combined with Medicare numbers and addresses, users face elevated risks of fraud. 	
		\item Loss of Trust: Erosion of public confidence in digital health infrastructure. 
	\end{itemize}
	
\textbf{Potential Consequences on Businesses:}

\begin{itemize}[leftmargin=*, itemsep=0pt, topsep=2pt]
	\item Significant financial losses: including legal action, customer compensation, data restoration, and security maintenance costs. 	
	\item Reputational damage: reduced customer trust in the company, leading to customer churn. 	
	\item Compliance and legal liability: potential violations of regulations such as the Privacy Act and GDPR, resulting in significant fines. 	
	\item Business interruption: Data breaches can lead to system downtime and operational delays, impacting profitability. 
\end{itemize}

\textbf{Regulatory References:}
\begin{itemize}[leftmargin=*, itemsep=0pt, topsep=2pt]
	\item GDPR Article 32: Requires encryption and pseudonymisation of sensitive data \cite{gdpr32}. 
	\item Australian Privacy Principles (APP 11.1): Mandates reasonable steps to protect personal information \cite{oaicAPP11}. 
	\item  Section 1798.100: Grants consumers rights to know, delete, and opt out of data collection \cite{ccpa1798}. 
\end{itemize}

Medibank’s failure to encrypt high‑entropy medical records, reliance on centralized storage, and lack of user control directly contravened these regulatory requirements, underscoring its systemic compliance gap.

\textbf{Data Entropy Analysis:}

\begin{itemize}[leftmargin=*, itemsep=0pt, topsep=2pt]
	\item Diagnosis codes and treatment types exhibit high entropy, meaning they contain significant information and pose a re-identification risk. Detailed data types, information entropy levels, and risk analysis in this data breach are shown in Table \ref{tab:entropy}. 
\end{itemize}

\end{enumerate}

\subsection{Formal Definition and Parameterization of Differential Privacy}

DP provides a mathematically rigorous framework for protecting individual-level information in statistical analysis. By introducing controlled random noise to query outputs, DP ensures that the inclusion or exclusion of a single record does not significantly affect the overall result. This mechanism limits an adversary’s ability to infer private details, even when auxiliary datasets are available.

From an information-theoretic perspective, \textbf{data entropy (H)} represents the degree of uncertainty or randomness within a dataset, defined as:

\begin{equation}
	H(X) = - \sum_{x} p(x)\log_{2} p(x)
\end{equation}

A dataset with \textbf{high entropy}—such as free-text clinical notes or rare diagnostic codes—contains more unique and less predictable information, thus posing a greater re-identification risk. Conversely, \textbf{low-entropy data} (e.g., categorical or aggregated attributes) exhibits higher redundancy and lower privacy risk.

By integrating entropy analysis with DP, privacy budgets can be \textbf{adaptively allocated}: fields or records with higher entropy receive stronger noise injection, achieving a better balance between \textbf{data utility and privacy protection}. In healthcare analytics, this entropy-aware DP design enables safer release of statistical summaries while preserving meaningful clinical patterns.

In the \textbf{Medibank breach}, such an entropy-aware DP mechanism was absent. Sensitive high-entropy attributes—including unstructured diagnostic descriptions and combined identity fields—were exposed in near-plaintext form, allowing attackers to re-associate fragmented records with real individuals. If DP had been applied with entropy-based budget calibration, these high-risk attributes could have been sufficiently perturbed to prevent re-identification while maintaining the analytical value of the dataset.

\noindent \textbf{1) Formal Definition of Differential Privacy}

A randomized algorithm $M$ satisfies \textit{($\varepsilon$, $\delta$)-DP} if, for any two adjacent datasets $D_1$ and $D_2$ that differ by at most one record, and for any possible output subset $S \subseteq \text{Range}(M)$:

\[
\Pr[M(D_1) \in S] \le e^{\varepsilon} \Pr[M(D_2) \in S] + \delta
\]

Here,  
$\varepsilon$ (epsilon) controls the privacy loss bound—smaller $\varepsilon$ implies stronger privacy.  
$\delta$ allows a small probability of this bound being violated; when $\delta = 0$, the mechanism satisfies \textit{pure differential privacy}.

\noindent \textbf{Typical mechanisms include:}

\begin{itemize}
	\item \textbf{Laplace Mechanism (for numerical queries):}
	\[
	M(D) = f(D) + \text{Lap}\!\left(\frac{\Delta f}{\varepsilon}\right)
	\]
	where $\Delta f$ is the $\ell_1$-sensitivity of the function $f$.
	
	\item \textbf{Gaussian Mechanism (for approximate DP):}
	\[
	M(D) = f(D) + \mathcal{N}(0, \sigma^2), \quad 
	\sigma \ge \frac{\sqrt{2 \ln(1.25 / \delta)} \, \Delta f}{\varepsilon}
	\]
	
	\item \textbf{Exponential Mechanism (for categorical outputs):}
	assigns selection probability proportional to
	\[
	\Pr[r] \propto \exp\!\left( \frac{\varepsilon \, u(D, r)}{2 \Delta u} \right),
	\]
	where $u(D, r)$ is a utility function.
\end{itemize}

\noindent \textbf{2) Privacy Budget Allocation}

In real-world systems such as Medibank’s analytics platform, multiple queries or models share a finite privacy budget $\varepsilon_{\text{total}}$.  
According to the \textit{sequential composition theorem}, the cumulative privacy loss satisfies:

\[
\varepsilon_{\text{total}} = \sum_{i=1}^{k} \varepsilon_i
\]

Thus, privacy budgets must be distributed adaptively across queries based on their sensitivity or entropy level.  
High-entropy attributes (e.g., free-text diagnoses, genomic codes) receive smaller $\varepsilon$ (i.e., more noise), while low-entropy aggregated data can tolerate larger $\varepsilon$ to preserve analytical accuracy.  
This entropy-aware budget allocation ensures that privacy protection intensity is proportional to re-identification risk, optimizing the trade-off between privacy and data utility.

\noindent \textbf{3) Parameter Selection Principles}

\begin{itemize}
	\item \textbf{Epsilon ($\varepsilon$):}  
	Typical values range from $0.01$ (strong privacy) to $5$ (weak privacy).  
	In regulated domains such as healthcare, $\varepsilon \approx 0.1$–$1.0$ is commonly adopted to balance privacy and analytical validity.  
	In adaptive DP systems, $\varepsilon$ can vary by data type or query sensitivity.
	
	\item \textbf{Delta ($\delta$):}  
	Represents the probability that DP fails to hold.  
	It should be smaller than the inverse of dataset size, i.e., $\delta < 1/|D|$.  
	For datasets with millions of records, $\delta \approx 10^{-6}$ is a standard safe choice.
\end{itemize}

\begin{table}[htbp]
	\caption{Data Entropy Analysis}
	\label{tab:entropy}
	\centering
	\renewcommand{\arraystretch}{1.2} % 调整行距
	\setlength{\tabcolsep}{8pt}       % 调整列间距
	\begin{tabular}{|c|c|c|}
		\hline
		\textbf{Field} & \textbf{Entropy Level} & \textbf{Risk Level} \\ \hline
		Name & Medium & Moderate \\ \hline
		\textbf{Medicare Number} & \textit{High} & \textbf{\textit{Severe}} \\ \hline
		\textbf{Diagnosis Code} & \textit{Very High} & \textbf{\textit{Critical}} \\ \hline
		\textbf{Treatment Type} & \textit{High} & \textbf{\textit{Severe}} \\ \hline
		Email Address & Medium & Moderate \\ \hline
	\end{tabular}
\end{table}

\subsection{Critical Appraisal}
\textbf{Critique of Existing Methods:}

\textbf{Encryption Gaps}: Although passwords were hashed, medical records were often stored in plaintext or in weakly protected formats. 

\textbf{Reactive Security Posture}: Medibank lacked proactive intrusion detection and zero-trust architecture, allowing lateral movement post-breach. 

\textbf{Absence of Multi-Factor Authentication (MFA)}: Attackers were able to infiltrate the system using compromised third-party credentials due to the lack of multi-factor authentication protocols \cite{cyberarrow2025}. 

Firewall Misconfiguration: The system did not enforce digital certificate validation for remote access, resulting in insufficient authentication and exposure to unauthorised entry \cite{siganto2024}. 

Lack of Data Tiering and Segmentation: Highly sensitive medical records were stored alongside general identity information within a unified system, without entropy-based classification or layered access controls. 

No Implementation of DP Mechanisms: Without good differential privacy protection for data, statistical results can easily expose too much information. Attackers can use re-identification attacks to associate diagnostic information, age, bills, and other combinations with specific individuals, further infer and obtain sensitive medical information, and cause more serious data breaches. 

\text{Shortcomings Summary:}

\begin{itemize}[leftmargin=2em, itemsep=0pt, topsep=2pt]
	\item Technical Deficiency: No noise-injection or privacy-preserving computation. 	
	\item Architectural Flaws: Centralised storage without segmentation. 	
	\item Regulatory Non-Compliance: Breach of APP 11.1 and GDPR Article 32.
\end{itemize}

\text{Proposed Solution Overview:}

To address these gaps, we propose a hybrid solution combining:

\begin{itemize}[leftmargin=2em, itemsep=0pt, topsep=2pt]
	\item Differential Privacy for statistical analytics 
	\item Data Entropy–based field classification
	\item Python-based simulation of privacy-preserving data release 
\end{itemize}

\text{Solution Justification:}

Why DP Is Best for Medibank:
Table\ref{tab:privacy-techniques} illustrates the differences between HE, SMPC, and DP. While HE and SMPC theoretically offer stronger "computational confidentiality" protection, they typically require extremely high computing resources, communication bandwidth, and protocol complexity, making them unsuitable for centralized healthcare data platforms like Medibank that rely on existing systems. In contrast, DP offers advantages such as lightweight, high flexibility, and mathematical verifiability, making it more suitable for direct integration into existing data analysis processes.

DP operates at the data analysis layer, introducing random noise into statistical queries or aggregated reports without modifying the underlying system structure, effectively preventing attackers from inferring individual information from the results. For Medibank, the key data breach issue wasn't cryptographic failure, but rather the exposure of sensitive statistical information in the analysis interface. Therefore, DP provides protection directly at the point of risk.

Furthermore, DP's design aligns closely with the "risk-based protection and data minimization" principles of APP 11.1 and Article 32 of the GDPR\cite{oaicAPP11}. Unlike HE and SMPC, which primarily prevent leakage during computation or transmission, DP effectively addresses the risk of post-processing leakage, as exposed in the Medibank incident—the indirect re-identification of high-entropy statistical data.

\section{Innovative Solution Design}

The three privacy anomalies identified in Section I.F require coordinated technical interventions:

\textbf{Unencrypted Storage of High-Entropy Medical Data} can be addressed by applying differential privacy to published statistical outputs, with entropy-aware noise calibration ensuring 
$(\varepsilon, \delta)$-DP guarantees.

\textbf{Lack of Data Segmentation and Access Control} can be resolved by establishing a zero-trust architecture with database segmentation, role-based privilege enforcement, and system-level access isolation.

\textbf{Absence of Privacy-Preserving Analytics} can be mitigated by integrating DP mechanisms into the analytical pipeline, ensuring that internal research and model training operate on noise-protected aggregates rather than raw records.

To operationalize these principles, we have designed the following five-layer defence framework:

\subsection{Database Layer}
At the database level, we anticipate protecting data at rest through structural segmentation, access isolation, and database hardening. Sensitive medical and personal information will be partitioned into multiple logical tables based on sensitivity and business function, minimising the risk of exposure to high-risk data. Role-based account controls will be implemented to enforce the principle of least privilege, ensuring that each department has access only to the data required for their mission. Furthermore, we intend to further protect stored information from unauthorised access or misuse through system-level protections such as AppArmor\footnote{AppArmor is a Linux security module that enforces mandatory access control policies on programs. It restricts the capabilities of applications by defining per-program profiles, thereby limiting potential damage from compromised or misbehaving software. AppArmor enhances system security by confining processes to a minimal set of required privileges\cite{apparmor_doc}.} to limit access scope, encrypted backups using mysqldump\footnote{mysqldump is a command-line utility provided by MySQL for exporting database contents into a logical backup format. It generates SQL statements that can recreate the database schema and data, facilitating migration, archival, and disaster recovery. The tool supports selective dumping of tables, databases, and advanced options for consistency and compression\cite{mysql_mysqldump}.} and GPG  (GNU Privacy Guard)\footnote{GPG is an open-source implementation of the OpenPGP standard that enables secure communication through encryption and digital signatures. It allows users to encrypt files, emails, and messages, verify authenticity, and manage cryptographic keys. GPG supports both symmetric and asymmetric encryption, ensuring confidentiality and integrity in data exchange\cite{openpgp_standard}.}, and audit logging. 

\subsubsection{Data Segmentation}
Divide the database into multiple logical tables, grouping them according to sensitivity and business function. Table \ref{tab:db-seg} shows the specific content.

\begin{table}[htbp]
	\caption{Database Segmentation}
	\label{tab:db-seg}
	\centering
	\begin{threeparttable}
		\renewcommand{\arraystretch}{1.15} % 行距稍增，易读
		\setlength{\tabcolsep}{5pt}        % 列间距
		\begin{tabular}{|>{\centering\arraybackslash}m{2.0cm}|
				>{\centering\arraybackslash}m{4.2cm}|
				>{\centering\arraybackslash}m{1.6cm}|}
			\hline
			\textbf{Table Name} & \textbf{Description} & \textbf{Sensitivity} \\ \hline
			
			patients & Basic customer information (Name, DOB, Address) & High \\ \hline
			medical\_records & Diagnosis, treatment, and medication details & High \\ \hline
			claims & Insurance claims records & Medium \\ \hline
			billing & Billing and payment information & Medium \\ \hline
			staff & Employee information and access rights & High \\ \hline
			audit\_logs & All access and operation logs & Medium \\ \hline
			
		\end{tabular}
	\end{threeparttable}
\end{table}

\subsubsection{Account Division and Authority Control}
Adopt the principle of least privilege and establish access rights separation via account grouping. Each department is restricted to access only the data tables they need. The specific separation details are shown in Table \ref{tab:acct-auth}. 

\begin{table}[htbp]
	\caption{Account Division and Authority Control}
	\label{tab:acct-auth}
	\centering
	\begin{threeparttable}
		\renewcommand{\arraystretch}{1.15}   % 行距
		\setlength{\tabcolsep}{3pt}          % 列间距（越小越省版面）
		\footnotesize                        % 表内字号（必要时可换 \scriptsize）
		\begin{tabular}{|>{\centering\arraybackslash}m{1.5cm}|
				>{\centering\arraybackslash}m{2.0cm}|
				>{\centering\arraybackslash}m{2.5cm}|
				>{\centering\arraybackslash}m{2.0cm}|}
			\hline
			\textbf{Username} & \textbf{Department} & \textbf{Privileges} & \textbf{Description} \\ \hline
			
			admin\_root & IT Security &
			ALL PRIVILEGES &
			Restrict to root administrator \\ \hline
			
			med\_ops & Medical Operations &
			SELECT, INSERT\newline on patients,\newline medical\_records &
			Medical data entry \\ \hline
			
			claims\_team & Insurance Claims &
			SELECT, INSERT,\newline UPDATE on claims &
			Claims processing \\ \hline
			
			billing\_team & Finance &
			SELECT, INSERT,\newline UPDATE on billing &
			Billing processing \\ \hline
			
			auditor & Compliance Audit &
			SELECT on all tables &
			Read-only audit access \\ \hline
			
			dev\_readonly & Tech Development &
			SELECT on patients,\newline claims &
			Limited to non-sensitive field testing \\ \hline
			
			backup\_agent & Operations &
			LOCK TABLES,\newline SELECT &
			For encrypted backup scripts \\ \hline
			
		\end{tabular}
	\end{threeparttable}
\end{table}

\subsubsection{Enhancement of Database Security}
To enhance the overall resilience of our data storage infrastructure, we have implemented the following configuration and access control measures to strengthen our database environment. 

\begin{itemize}[leftmargin=2em, itemsep=2pt, topsep=2pt]
	\item Enable the MySQL audit log plugin to record all queries and administrative operations.
	
	\item Enable mandatory encrypted connections (SSL/TLS) to encrypt communications between the client and the database, preventing sensitive data from being intercepted or exploited in man-in-the-middle attacks.
	
	\item Disable remote root logins. The root user is only allowed locally; remote access is prohibited.
	
	\item Enable MFA for database logins. Direct login with credentials alone is not possible.
	
	\item Enable jump host logins. The database cannot be accessed remotely. All access must go through a configured jump host. Ensure that the database is physically and logically isolated from external networks.
\end{itemize}

\subsubsection{Using mysqldump and GPG to Back Up Database}
Use mysqldump to export database contents in plain text, and GPG to encrypt the exported plain text, ensuring that the backup cannot be directly read during storage and transmission.

\subsection{Network Layer}
We plan to combine the deployment of a firewall and a virtual private network (VPN) to effectively prevent eavesdropping, session hijacking, and unauthorised access, further strengthening overall network security. 

\subsubsection{Firewall}
Divide the network into multiple security zones. Strictly restrict access paths between zones using firewall policies. Use port whitelisting and protocol detection to limit inbound and outbound traffic. Role-based policies further restrict access and adhere to the principle of least privilege, ensuring that different departmental roles, such as medical operators, claims teams, and auditors, can access only the internal assets necessary for their tasks. This effectively reduces the attack surface and prevents lateral movement of intruders within the healthcare network.  

\subsubsection{VPN}
Set up a VPN to secure remote access to the company's internal network. Enable both credentials and MFA. Only authenticated users with valid credentials and MFA can connect remotely. All external links must go through the VPN tunnel, using robust encryption protocols like IPSec to protect data confidentiality and integrity during transit. The VPN gateway must verify user identities and keep records of all access, creating an audit trail for regulatory compliance.

\subsection{System Layer}

\subsubsection{Use AppArmor to Restrict MySQL User Access}
AppArmor ensures that MySQL processes can only access files and directories related to database operations, preventing them from freely accessing other system resources. Even if an attacker successfully exploits a vulnerability to gain access to the database process, they cannot use it to read, modify, or steal non-database files, thus reducing the potential attack surface.  

\subsubsection{System Account and Access Control}
Table \ref{tab:sys-accounts} lists the basic information for each system account on the server, including the account name, department, and primary responsibilities within the system, and clarifies the business affiliation and functional positioning of each account, providing a basis for subsequent permission configuration. 

\begin{table}[htbp]
	\caption{System Accounts and Responsibilities}
	\label{tab:sys-accounts}
	\centering
	\begin{threeparttable}
		\renewcommand{\arraystretch}{1.15}   % 行距
		\setlength{\tabcolsep}{4pt}          % 列间距
		\footnotesize                        % 表内字号
		\begin{tabular}{|>{\centering\arraybackslash}m{1.3cm}|
				>{\centering\arraybackslash}m{2.0cm}|
				>{\centering\arraybackslash}m{4.7cm}|}
			\hline
			\textbf{Username} & \textbf{Department} & \textbf{Main Responsibility} \\ \hline
			
			medops & Medical Ops & Handle patient data upload/import \\ \hline
			claims & Claims & Upload claim docs and trigger scripts \\ \hline
			billing & Finance & Import billing data, generate payments \\ \hline
			auditor & Audit & Read audit logs and backup snapshots \\ \hline
			dev & Development & Run test scripts, generate mock data \\ \hline
			backup & Operations & Automated DB and file backups \\ \hline
			reportbot & Data Analysis & Generate and export reports \\ \hline
			mysql & System Service & Run MySQL service \\ \hline
			admin & IT Security & Manage all accounts and privileges \\ \hline
			
		\end{tabular}
	\end{threeparttable}
\end{table}

Table \ref{tab:access-perm} details the directory paths accessible by each account on the server and the corresponding permission types (such as read, write, and execute). It defines the access scopes of different roles, ensuring that each account adheres to the principle of least privilege. 

\begin{table}[htbp]
	\caption{Accessible Directories and Permissions}
	\label{tab:access-perm}
	\centering
	\begin{threeparttable}
		\renewcommand{\arraystretch}{1.15}   % 行距
		\setlength{\tabcolsep}{3pt}          % 列间距
		\footnotesize                        % 表内字号
		\begin{tabular}{|>{\centering\arraybackslash}m{1.8cm}|
				>{\centering\arraybackslash}m{3.8cm}|
				>{\centering\arraybackslash}m{2.6cm}|}  % 总宽度约 8cm
			\hline
			\textbf{Username} & \textbf{Accessible Directory} & \textbf{Permission Type} \\ \hline
			
			medops & /data/patients/ & Read/Write \\ \hline
			claims & /data/claims/ & Read/Write \\ \hline
			billing & /data/billing/ & Read/Write \\ \hline
			auditor & /var/log/mysql/\newline /secure\_backups/ & Read-Only \\ \hline
			dev & /home/dev/\newline /data/sandbox/ & Read/Write/Execute \\ \hline
			backup & /secure\_backups/\newline /etc/mysql/ssl/ & Read/Write/Execute \\ \hline
			reportbot & /reports/\newline /secure\_backups/ & Write \\ \hline
			mysql & /var/lib/mysql/\newline /etc/mysql/ & System Account \\ \hline
			admin & All directories & Full RWX \\ \hline
			
		\end{tabular}
	\end{threeparttable}
\end{table}

Table \ref{tab:special-account} lists each account's special system privileges (such as sudo permissions) and corresponding security recommendations. By clarifying security control measures (such as disabling the shell, enabling SFTP, using SSH keys, and enabling 2FA), we can strengthen overall system security and prevent unauthorised operations and security vulnerabilities. 

\begin{table}[htbp]
	\caption{Special Account Permissions and Security Recommendations}
	\label{tab:special-account}
	\centering
	\begin{threeparttable}
		\renewcommand{\arraystretch}{1.15}   % 行距
		\setlength{\tabcolsep}{3pt}          % 列间距
		\footnotesize                        % 表内字号
		\begin{tabular}{|>{\centering\arraybackslash}m{1.6cm}|
				>{\centering\arraybackslash}m{2.8cm}|
				>{\centering\arraybackslash}m{3.8cm}|}  % 总宽度约 8cm
			\hline
			\textbf{Username} & \textbf{Special Privilege} & \textbf{Security Recommendation} \\ \hline
			
			medops & No sudo & Disable shell, allow SFTP only \\ \hline
			claims & No sudo & Restrict IP login \\ \hline
			billing & No sudo & Restrict group privileges \\ \hline
			auditor & No sudo & Enforce SSH key use \\ \hline
			dev & No sudo & Restrict production data access \\ \hline
			backup & sudo mysqldump, gpg & Join backup group,\newline enable AppArmor \\ \hline
			reportbot & No sudo & Disable shell, allow group only \\ \hline
			mysql & System account & Restricted by AppArmor \\ \hline
			admin & Full sudo & Allow jump server login only,\newline enable 2FA \\ \hline
			
		\end{tabular}
	\end{threeparttable}
\end{table}

Table \ref{tab:account-groups} describes the business responsibilities of different account groups in the system and their typical member accounts. Each account group represents a functional department (such as healthcare operations, claims, finance, audit, etc.), and its member accounts are responsible for corresponding operational tasks. 

\begin{table}[htbp]
	\caption{System Account Groups and Typical Accounts}
	\label{tab:account-groups}
	\centering
	\begin{threeparttable}
		\renewcommand{\arraystretch}{1.15}
		\setlength{\tabcolsep}{3pt}
		\footnotesize
		\begin{tabular}{|>{\centering\arraybackslash}m{1.6cm}|
				>{\centering\arraybackslash}m{4.6cm}|
				>{\centering\arraybackslash}m{2.0cm}|}
			\hline
			\textbf{Group} & \textbf{Description} & \textbf{Typical Member Account} \\ \hline
			medops & Medical Ops Group: Handle patient data and medical records & medops \\ \hline
			claims & Claims Group: Process insurance claim documents and workflows & claims \\ \hline
			billing & Billing Group: Manage billing and payment data & billing \\ \hline
			audit & Audit Group: Read logs and backup snapshots & auditor \\ \hline
			dev & Development Group: Run test scripts and generate data & dev \\ \hline
			backup & Backup Group: Perform encrypted backups and recovery & backup \\ \hline
			report & Report Group: Generate and export periodic reports & reportbot \\ \hline
			mysql & MySQL Service Group: Run database service & mysql \\ \hline
			admin & Admin Group: Manage all system accounts and permissions & admin \\ \hline
			
		\end{tabular}
	\end{threeparttable}
\end{table}

Table \ref{tab:group-access} defines directory access permissions and permission settings for each system group, outlining the directories they can read, write, or execute. Each group (for example, Healthcare Operations, Audit, and Development) is assigned specific access levels based on their business role. Restrictions and security controls (for example, limiting access to medical records, using sudo permissions, or using jump server logins) are implemented to enforce the principle of least privilege and ensure secure data processing across the system.

\begin{table}[htbp]
	\caption{Group Access Directory and Permissions}
	\label{tab:group-access}
	\centering
	\begin{threeparttable}
		\renewcommand{\arraystretch}{1.15}   % 行距
		\setlength{\tabcolsep}{2.5pt}        % 列间距
		\footnotesize                        % 表内字号
		\begin{tabular}{|>{\centering\arraybackslash}m{1.0cm}|
				>{\centering\arraybackslash}m{2.2cm}|
				>{\centering\arraybackslash}m{2.0cm}|
				>{\centering\arraybackslash}m{2.8cm}|}  % 总宽度约 8cm
			\hline
			\textbf{Group} & \textbf{Access Directory} & \textbf{Permission Type} & \textbf{Notes} \\ \hline
			
			medops & /data/patients/ & Read/Write & Restricted from claims/billing \\ \hline
			claims & /data/claims/ & Read/Write & No access to medical\_records \\ \hline
			billing & /data/billing/ & Read/Write & No access to patient data \\ \hline
			audit & /var/log/mysql/\newline /secure\_backups/ & Read-Only & Cannot write any data \\ \hline
			dev & /home/dev/\newline /data/sandbox/ & Read/Write\newline/Execute & No access to production data \\ \hline
			backup & /secure\_backups/\newline /etc/mysql/ssl/ & Read/Write\newline/Execute & sudo mysqldump, gpg allowed \\ \hline
			report & /reports/\newline /secure\_backups/ & Write & Shell disabled, SFTP only \\ \hline
			mysql & /var/lib/mysql/\newline /etc/mysql/ & Read/Write & Default system service account \\ \hline
			admin & All directories & Full RWX & Full sudo, jump server login only \\ \hline
			
		\end{tabular}
	\end{threeparttable}
\end{table}

Table \ref{tab:key-dir} lists the paths, purposes, accessible accounts, and corresponding permissions for each key directory in the server system. It clarifies the functional positioning and access scope of different directories, helping administrators understand which accounts can read, write, or execute specific data. 

\begin{table}[H]
	\caption{Key Directory Description}
	\label{tab:key-dir}
	\centering
	\begin{threeparttable}
		\renewcommand{\arraystretch}{1.15}
		\setlength{\tabcolsep}{2.5pt}
		\footnotesize
		\begin{tabular}{|>{\centering\arraybackslash}m{2.0cm}|
				>{\centering\arraybackslash}m{2.6cm}|
				>{\centering\arraybackslash}m{1.8cm}|
				>{\centering\arraybackslash}m{1.8cm}|}
			\hline
			\textbf{Directory Path} & \textbf{Description} & \textbf{Accessible Accounts} & \textbf{Permission Type} \\ \hline
			
			/data/patients/ & Directory for patient data uploads & medops, admin & Read/Write \\ \hline
			/data/claims/ & Directory for insurance claim documents & claims, admin & Read/Write \\ \hline
			/data/billing/ & Directory for billing and payment data & billing, admin & Read/Write \\ \hline
			/secure\_backups/ & Encrypted backup directory & backup, auditor, admin & Read/Write (partially Read-Only) \\ \hline
			/var/log/mysql/ & MySQL audit log directory & auditor, admin & Read-Only \\ \hline
			/etc/mysql/ssl/ & SSL certificate directory & backup, admin & Read-Only \\ \hline
			/reports/ & Report export directory & reportbot, admin & Write \\ \hline
			/home/dev/ & Development test directory & dev, admin & Read/Write\newline/Execute \\ \hline
			
		\end{tabular}
	\end{threeparttable}
\end{table}

\subsection{Algorithmic Layer}
We employ six different differential privacy mechanisms for various data attributes, aiming to strike a balance between the privacy and availability of each attribute. 

\subsubsection{Laplace Mechanism}\footnote{The Laplace mechanism achieves differential privacy by injecting noise drawn from the Laplace distribution into query results. This noise is calibrated to the query's sensitivity, ensuring that individual data points have minimal influence on the output. It is especially effective for numeric queries with bounded sensitivity, offering a straightforward way to balance privacy and utility\cite{google_privacy_utility_tradeoff}.}

Applied to the diagnosis code field (diagnosis counts). It adds noise to the number of patients in each diagnosis category, thereby masking the contribution of individual patients. This makes it difficult for an attacker to determine whether an individual has the disease based on discrepancies in the counts.
 
\subsubsection{Gaussian Mechanism}\footnote{The Gaussian mechanism introduces normally distributed noise to query outputs, tailored to both the sensitivity of the query and the desired privacy parameters. It is particularly suited for approximate differential privacy $(\varepsilon, \delta)$, where a small probability of privacy breach is acceptable. This method is widely used in machine learning applications due to its favorable error bounds in high-dimensional settings\cite{google_dp_gaussian}.} 

Applied to the Age field (calculating the mean age). By adding Gaussian noise to the mean, the result complies with the $(\varepsilon, \delta)$-DP criterion. This ensures that even if the ages of a few patients are modified or leaked, the overall mean age result does not significantly reveal information about any individual.

\subsubsection{Exponential Mechanism}\footnote{The exponential mechanism selects outputs from a set of possible results based on a utility function, favoring those with higher utility while preserving privacy. Instead of adding noise to numeric values, it probabilistically chooses outcomes, making it ideal for non-numeric tasks such as classification or ranking. Its design ensures that the most useful outputs remain likely even under strict privacy constraints\cite{google_optimal_noise}.} 

Applied to the treatment type field (selecting treatment types). It assigns selection probabilities based on a utility function to select a treatment type from multiple candidate types. The exponential mechanism introduces randomness into the selection process (based on the utility function and the privacy budget $\varepsilon$), preventing attackers from inferring the true data distribution from the selection results.

\subsubsection{Randomised Response}\footnote{Randomised response is a foundational technique in local differential privacy, allowing individuals to respond to sensitive queries with randomized answers. This method ensures plausible deniability for participants while enabling accurate aggregate statistics. It is especially useful in surveys and decentralized data collection where trust in a central curator is limited\cite{google_regression_label_dp}.}

Applied to the abortion flag field (a binary sensitive attribute). When a patient answers the question "whether they have had an abortion," each response is flipped with a certain probability, thus providing local differential privacy. This function protects patient privacy on the input side, allowing researchers to estimate only the overall proportions and not confirm the true answers for specific individuals.

\subsubsection{Histogram Mechanism}\footnote{The histogram mechanism privately estimates frequency distributions by adding calibrated noise to each bin count. It enables the release of categorical data summaries while preserving individual privacy. Advanced variants, such as those in the shuffled model, improve accuracy by reducing domain-size dependence and enhancing sample efficiency\cite{harvard_histogram_dp}.}

Applied to the treatment type field (the distribution of the number of people in different treatment categories). By independently adding noise to the counts of each category, a differentially private histogram is generated. This allows researchers to analyse the overall distribution trends of treatment options while reducing sensitivity to individual patient choices.

\subsubsection{Sparse Vector Technique}\footnote{The sparse vector technique allows repeated queries against a private dataset, revealing only those that exceed a threshold while maintaining privacy. It efficiently manages the privacy budget by limiting the number of noisy outputs, making it suitable for monitoring tasks or adaptive query answering. Recent refinements improve its robustness and applicability in dynamic settings\cite{google_sparse_vector}.}

Applied to the Age field (threshold queries). This mechanism allows publishing only a few results that exceed a threshold in a large number of queries, while adding noise to the remaining results. This ensures that the overall privacy budget is not quickly exhausted when executing multiple queries, allowing efficient answers to threshold questions such as "How many patients are over 65?"

\subsection{Enterprise Layer}
Enterprise Security Suite:
\begin{itemize}[leftmargin=2em, itemsep=4pt, topsep=2pt]
	
	\item Enable a Zero Trust architecture. No network traffic is trusted; every access request requires authentication.
	
	\item MFA must be enabled for all critical systems and internal accounts.
	
	\item Enable access control. Use role-based access control (RBAC) and attribute-based access control (ABAC) policies to strictly limit access to only necessary permissions.
	
	\item Deploy an intrusion detection system (IDS) and an intrusion prevention system (IPS) to monitor and block abnormal traffic and promptly identify potential attacks.
	
	\item Periodically conduct security audits to regularly review configurations, permissions, and data access logs to identify potential vulnerabilities.
	
	\item Periodically conduct employee security awareness training to help employees identify non-technical threats such as phishing emails and social engineering attacks, enhance their security awareness, and reduce the potential for human error.
	
\end{itemize}

\section{Hypothetical Deployment}
Since we don't have a firewall device, we will use a jump host to configure the OpenVPN\footnote{OpenVPN is a versatile VPN solution that establishes encrypted connections using SSL/TLS protocols. It enables secure communication across untrusted networks by creating virtual tunnels that protect data in transit. With support for multiple authentication methods and encryption standards, OpenVPN is widely used for remote access, site-to-site connectivity, and enterprise-grade security deployments. Its modular configuration and cross-platform compatibility make it suitable for both lightweight clients and complex network infrastructures\cite{openvpn_tls}.} service as the simulation solution and access the MySQL database server only through the jump host. We use Ubuntu 22.04 as the jump host. In addition, we will enable OpenVPN's logging function to facilitate regular security audits.

This chapter describes a concrete, reproducible deployment plan that implements the layered solution from Chapter II on an Ubuntu 24.04 host. It covers selected software versions, the test dataset, precise implementation steps for database segmentation, system hardening, backup and key management, AppArmor profile deployment, network controls (jump host + VPN), and the differential privacy processing pipeline implemented in Python.

\begin{figure}[htbp]
	\centering
	\includegraphics[width=0.45\textwidth]{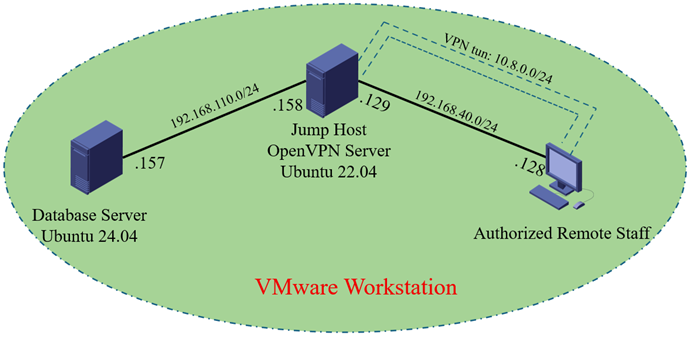}
	\caption{Network Logical Topology Diagram}
	\label{fig:sltd}
\end{figure}

Figure~\ref{fig:sltd} illustrates the logical topology of the experimental deployment within VMware Workstation.
The setup includes a Database Server (192.168.110.157, Ubuntu 24.04), an OpenVPN Jump Host (192.168.40.158, Ubuntu 22.04), and an Authorised Remote Client (192.168.40.128).
The jump host bridges two virtual networks (192.168.110.0/24 and 192.168.40.0/24) and provides VPN access over the 10.8.0.0/24 tunnel.
This architecture simulates a realistic segmented infrastructure where sensitive database systems are isolated and accessible only through controlled, encrypted VPN connections.

\subsection{Deployment Environment and Version Selection}
Our solution is deployed on Ubuntu 24.04 LTS, chosen for its long-term support and robust security patching. The database engine is MySQL 8.0.36, selected for its mature role-based access control, native audit logging, and SSL/TLS support. We use Python 3.11 for its performance and compatibility with modern privacy libraries, including OpenDP\footnote{OpenDP is a modular, open-source library developed to support rigorous differential privacy implementations in Python. It provides a suite of validated mechanisms, transformations, and privacy accounting tools that enable developers and researchers to build privacy-preserving data workflows. Designed with formal verification and reproducibility in mind, OpenDP emphasizes transparency and correctness, making it suitable for academic, governmental, and enterprise-grade privacy applications\cite{opendp_github}.} and diffprivlib\footnote{diffprivlib is a Python library created by IBM Research to integrate differential privacy into standard machine learning workflows. Built atop scikit-learn, it offers privacy-preserving versions of common models and preprocessing tools, allowing seamless substitution in existing pipelines. The library supports customizable privacy budgets and mechanisms, making it practical for real-world deployments where data utility and privacy must be balanced\cite{ibm_diffprivlib}.}, which support customizable differential privacy mechanisms.

To establish a role-based access model that meets security requirements, we implemented a strict access control system. The specific steps are as follows:

1) We created eight independent user accounts: \texttt{medops}, \texttt{claims}, \texttt{billing}, \texttt{auditor}, \texttt{dev}, \texttt{backup}, \texttt{reportbot}, and \texttt{admin}. The \texttt{dev} and \texttt{admin} users were configured to log in to the system, using \texttt{/bin/bash} as their login shell, allowing them to access the system command line and perform operations. The other six users (\texttt{medops}, \texttt{claims}, \texttt{billing}, \texttt{auditor}, \texttt{backup}, and \texttt{reportbot}) were configured as \texttt{/usr/sbin/nologin}, preventing them from logging in and only allowing them to run as system processes or access files.

2) We created eight groups corresponding to these users: \texttt{medops}, \texttt{claims}, \texttt{billing}, \texttt{audit}, \texttt{dev}, \texttt{backup}, \texttt{report}, and \texttt{admin}. Each group provides an access control unit for the corresponding user and similar tasks, facilitating the unified allocation of directory permissions.

3) We added each user to a group with the same name as theirs. For example, user \texttt{medops} is added to group \texttt{medops}, user \texttt{claims} is added to group \texttt{claims}, and so on. This step ensures that the system can identify each user's group membership when accessing files, thereby determining permissions.

4) We set the owner of each data directory to \texttt{root} and changed the group to the corresponding business group. We then set access permissions based on the directory's purpose:

\begin{itemize}
	\item The \texttt{/data/patients/} directory has group \texttt{medops} and permissions of \texttt{770}. Only the \texttt{root} user and members of the \texttt{medops} group can read, write, and execute; other users are completely blocked.
	\item The \texttt{/data/claims/} directory has group \texttt{claims} and permissions of \texttt{770}. Only \texttt{root} and members of the \texttt{claims} group can access it.
	\item The \texttt{/data/billing/} directory has group \texttt{billing} and permissions of \texttt{770}. Only \texttt{root} and members of the \texttt{billing} group can access it.
	\item The \texttt{/data/staff/} directory has group \texttt{staff} and permissions of \texttt{770}. Only \texttt{root} and members of the \texttt{staff} group can access it.
	\item The \texttt{/var/log/mysql/} directory has group permissions of \texttt{750}. \texttt{root} has read, write, and execute permissions. Members of the \texttt{audit} group only have read and execute permissions; other users have no access.
	\item The \texttt{/data/backups/} directory has group permissions of \texttt{750}. \texttt{root} has read, write, and execute permissions. Members of the \texttt{backup} group have read and execute permissions; no one else has access.
	\item The \texttt{/data/reports/} directory has group permissions of \texttt{750}. \texttt{root} has read, write, and execute permissions. Members of the \texttt{report} group have read and execute permissions; no one else has access.
\end{itemize}

After completing the above operations, the following goals can be achieved:
\begin{itemize}
	\item Each business department has a separate system user and user group.
	\item Access permissions for each directory are consistent with the group to which it belongs, preventing cross-department data access.
	\item The ultimate owner of all directories is \texttt{root}, ensuring administrative control.
	\item Only the \texttt{dev} and \texttt{admin} users can log in directly to the system and execute commands.
	\item Other service users (such as \texttt{reportbot} and \texttt{backup}) can only access specific directories through automated tasks or system programs.
\end{itemize}

Table~\ref{tab:patient-sample} shows the 8 rows of sample data we created.
\begin{table*}[htbp]
	\caption{Sample of Patient Records in Medibank Dataset}
	\label{tab:patient-sample}
	\centering
	\resizebox{\textwidth}{!}{%
		\begin{tabular}{|c|c|c|c|c|c|c|c|c|}
			\hline
			\textbf{ID} & \textbf{Name} & \textbf{Email} & \textbf{Date of Birth} & \textbf{Medicare Number} & \textbf{Diagnosis Code} & \textbf{Treatment Type} & \textbf{Address} & \textbf{Phone} \\ \hline
			1 & Allison Hill & donaldgarcia@example.net & 1946/8/4 & 1043321821 & Z71.3 & Speech Therapy & 133 Anna Trail, Robinsonshire, SA, 1265 & (03) 3511 6155 \\ \hline
			2 & Renee Blair & dudleynicholas@example.net & 1974/8/29 & 133898081 & J45.9 & Diabetes Education & 1 Donna Walkway, Traciebury, NSW, 2984 & +61.7.2553.4192 \\ \hline
			3 & Danielle Ford & veronica83@example.net & 1978/11/23 & 8637940299 & L40.0 & Specialist Referral & 564 Jason Ring, Jasonfort, VIC, 2902 & (03) 3884 9696 \\ \hline
			4 & Zachary Taylor & ddavis@example.org & 1955/7/24 & 4235116155 & F33.2 & Mental Health Counseling & Flat 66 7 Maddox Alleyway, New Kaylamouth, NSW, 2926 & 08-0482-8148 \\ \hline
			5 & Brittany Farmer & georgetracy@example.org & 1984/2/28 & 4078161847 & Z86.3 & Mental Health Counseling & 391 Jessica Bridge, West Donna, NT, 2789 & 61-3-8346-5787 \\ \hline
			6 & Danny Morgan & briannasmith@example.net & 1942/10/9 & 5931034139 & F41.1 & Physiotherapy & 51a Joshua Plaza, West Jennifer, WA, 2697 & 61-3-1165-6670 \\ \hline
			7 & Victoria Garcia & zchandler@example.org & 1968/8/7 & 4752558499 & Z86.3 & Surgical Procedure & 7 Robert Formation, Wrightland, WA, 9108 & 1326 7736 \\ \hline
			8 & Carmen Smith & ybaker@example.com & 1973/7/31 & 9288276491 & Z00.0 & Chemotherapy & Suite 343 980 Brown Riviera, Shawhaven, NT, 2281 & 02-9136-1939 \\ \hline
		\end{tabular}%
	}
\end{table*}

\begin{figure}[htbp]
	\centering
	\includegraphics[width=0.45\textwidth]{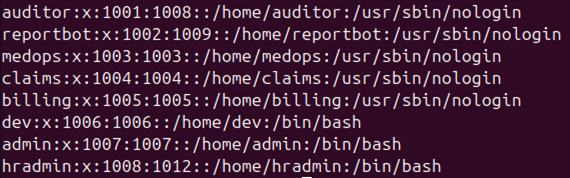}
	\caption{System User Creation Results}
	\label{fig:sucr}
\end{figure}

Figure~\ref{fig:sucr} presents the user account configuration within the Ubuntu operating system.
Each service account (e.g., auditor, reportbot, medops, claims, billing) is configured with /usr/sbin/nologin to prevent direct shell access, while administrative users such as dev and admin are assigned /bin/bash.
This design enforces strict privilege separation and ensures that only essential system operators can execute commands on the host environment.

\begin{figure}[htbp]
	\centering
	\includegraphics[width=0.45\textwidth]{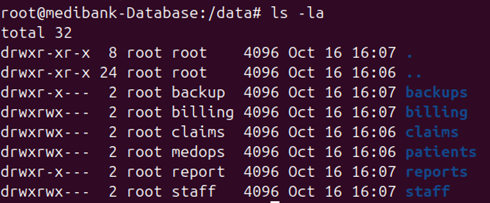}
	\caption{Directory Permission Settings}
	\label{fig:dps}
\end{figure}

Figure~\ref{fig:dps} shows the permission configuration of the \texttt{/data/} directory hierarchy, illustrating the mapping between business groups and their respective data directories. Each directory (e.g., \texttt{/data/patients/}, \texttt{/data/claims/}, \texttt{/data/billing/}) is owned by \texttt{root} and assigned group ownership corresponding to the relevant department. Access permissions are set to \texttt{770} or \texttt{750}, ensuring that only the appropriate group members and the root administrator can access sensitive data, effectively preventing cross-department data leakage.

\subsection{Database Segmentation and System Account Mapping}
\begin{figure*}[htbp]
	\centering
	\subfloat[audit\_logs Table Structure]{%
		\includegraphics[width=0.32\textwidth]{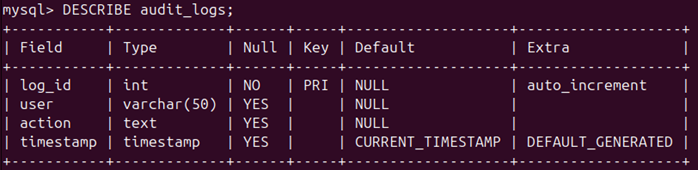}
	}\hfill
	\subfloat[staff Table Structure]{%
		\includegraphics[width=0.32\textwidth]{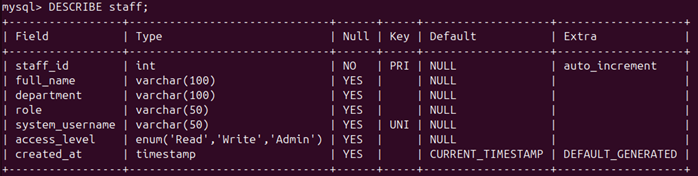}
	}\hfill
	\subfloat[billing Table Structure]{%
		\includegraphics[width=0.32\textwidth]{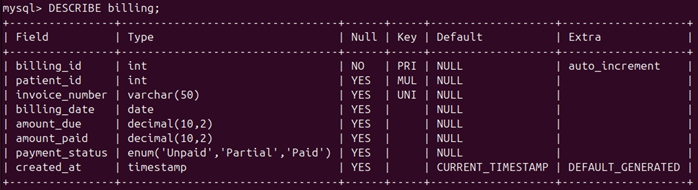}
	}\\
	\subfloat[claims Table Structure]{%
		\includegraphics[width=0.32\textwidth]{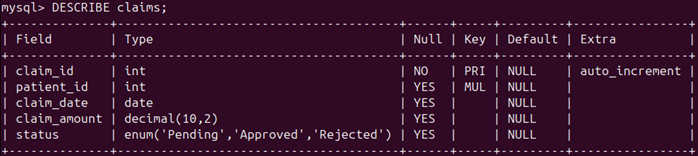}
	}\hfill
	\subfloat[medical\_records Table Structure]{%
		\includegraphics[width=0.32\textwidth]{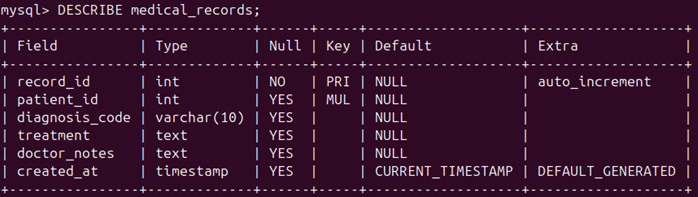}
	}\hfill
	\subfloat[patients Table Structure]{%
		\includegraphics[width=0.32\textwidth]{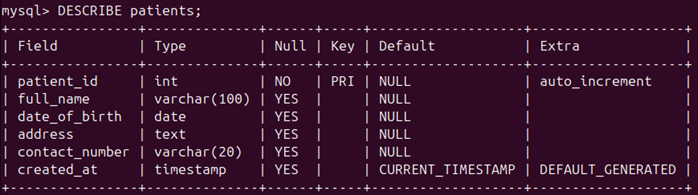}
	}
	\caption{Each Table Structure}
	\label{fig:eval-results}
\end{figure*}

Following the architectural design in Chapter~2, we implemented a segmented database architecture with five logical tables: \texttt{patients}, \texttt{medical\_records}, \texttt{claims}, \texttt{billing}, and \texttt{audit\_logs}. Each table is categorized by sensitivity level and mapped to specific system accounts and groups defined in Table~\ref{tab:db-seg}. The specific steps are as follows:

\begin{enumerate}
	\item We created a database named \texttt{medibank\_secure} to store medical business data. The character set was specified as \texttt{utf8mb4} and the collation as \texttt{utf8mb4\_unicode\_ci}, ensuring secure and compatible storage of multilingual characters (including Chinese, English, and symbols) while preventing character encoding issues.
	
	\item We created seven database accounts, each corresponding to a different system role or service:
	
	\begin{itemize}
		\item \texttt{admin\_root@localhost} \\
		Password: \texttt{StrongAdminPass!} \\
		Localhost-only login. Serves as the database administrator account, responsible for creating tables, assigning permissions, and maintaining the system.
		
		\item \texttt{med\_ops@\%} \\
		Password: \texttt{MedOpsPass!} \\
		Remote access permitted from any host. Corresponds to the “Healthcare Operations” function, responsible for patient and medical record data operations.
		
		\item \texttt{claims\_team@\%} \\
		Password: \texttt{ClaimsPass!} \\
		Remote access enabled for the claims processing department.
		
		\item \texttt{billing\_team@\%} \\
		Password: \texttt{BillingPass!} \\
		Remote access enabled for billing and payment-related data operations.
		
		\item \texttt{auditor@\%} \\
		Password: \texttt{AuditPass!} \\
		Remote access permitted for auditing and log review.
		
		\item \texttt{dev\_readonly@\%} \\
		Password: \texttt{DevReadPass!} \\
		Remote access permitted for developer debugging and read-only data review.
		
		\item \texttt{backup\_agent@localhost} \\
		Password: \texttt{BackupPass!} \\
		Localhost-only access, used for performing database backup tasks.
	\end{itemize}

	The user accounts and passwords we created are for demonstration purposes only. In real-world scenarios, longer and more complex passwords should be used, generated as randomly as possible to prevent social engineering attacks.

	Each user is assigned minimum permissions and distinct data access scopes to prevent accidental or unauthorized access.
	
	\item Figure \ref{fig:eval-results} shows that we established six core business tables in the database, covering patient information, medical records, claims, billing, logs, and employee management. Each table is associated with specific business operations and linked through relational keys to ensure data integrity and access control. Figure~\ref{fig:eval-results} shows the structure of these six main tables: \texttt{audit\_logs}, \texttt{patients}, \texttt{medical\_records}, \texttt{claims}, \texttt{staff}, and \texttt{billing}. Together, they store and manage all operational, medical, and administrative data within the system.
	
	\begin{itemize}
		\item \textbf{\texttt{patients}} — Stores basic patient information such as full name, date of birth, address, and contact number. Each record is uniquely identified by \texttt{patient\_id}, which serves as a primary key referenced by other tables.
		
		\item \textbf{\texttt{medical\_records}} — Records patients’ diagnoses and treatment details. The field \texttt{patient\_id} links each record to a patient in the \texttt{patients} table, maintaining one-to-one correspondence.
		
		\item \textbf{\texttt{claims}} — Tracks patients’ medical insurance claims, including claim date, amount, and approval status (\textit{Pending}, \textit{Approved}, or \textit{Rejected}). The field \texttt{patient\_id} associates each claim with a patient.
		
		\item \textbf{\texttt{billing}} — Manages financial and billing information such as invoice number, billing date, amount due, and payment status (\textit{Unpaid}, \textit{Partial}, or \textit{Paid}). Linked to \texttt{patients} via \texttt{patient\_id}.
		
		\item \textbf{\texttt{audit\_logs}} — Records user actions, including username, operation type, and timestamp, providing traceability for system audits and anomaly detection.
		
		\item \textbf{\texttt{staff}} — Contains employee details such as department, role, username, and access level (\textit{Read}, \textit{Write}, or \textit{Admin}), supporting access control and internal permission management.
	\end{itemize}
\end{enumerate}

\begin{figure*}[htbp]
	\centering
	\subfloat[Business and Support Users]{
		\includegraphics[width=0.32\textwidth]{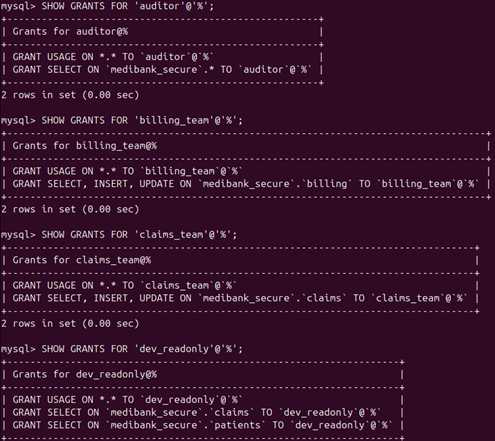}
	}\hfill
	\subfloat[\texttt{med\_ops} and \texttt{admin\_root} Users]{
		\includegraphics[width=0.64\textwidth]{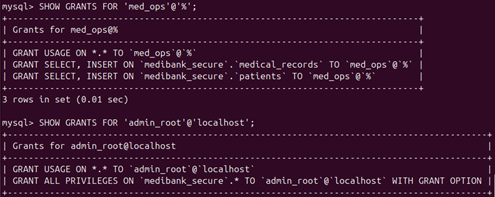}
	}
	\caption{Database Privilege Grants for Users}
	\label{fig:3&4}
\end{figure*}

Figures~\ref{fig:3&4} illustrate the MySQL privilege configuration for different database accounts in the \texttt{medibank\_secure} system. 
As shown in Fig.~\ref{fig:3&4}(b), the \texttt{med\_ops} account is granted limited access (\texttt{SELECT} and \texttt{INSERT}) to the \texttt{patients} and \texttt{medical\_records} tables, 
while the \texttt{admin\_root} account holds full administrative privileges with the \texttt{GRANT OPTION}, enabling complete control over all database objects. 
Fig.~\ref{fig:3&4}(a) further presents the access permissions of other functional and support accounts, including \texttt{auditor}, \texttt{billing\_team}, \texttt{claims\_team}, and \texttt{dev\_readonly}. 
Each account is restricted to specific tables and operations based on its business role, enforcing the principle of least privilege and minimising the potential impact of unauthorised access or accidental modification.

During the database backup and restoration process, the \texttt{GPG} decryption command is first executed to decrypt the encrypted backup file. 
The command syntax is as follows:

\begin{lstlisting}[language=bash]
	gpg --output medibank_backup.sql \
	--decrypt medibank_backup.sql.gpg
\end{lstlisting}

After the command is executed, the system prompts the user to enter a decryption passphrase. 
Only when the correct passphrase is provided can the backup file be successfully decrypted and restored. 
If an incorrect passphrase is entered, the system displays the message 
\texttt{"decryption failed: Bad session key"}, indicating that the decryption operation has failed and the file remains encrypted. 
This behaviour ensures that encrypted backup files cannot be accessed or tampered with when an invalid passphrase is used, thereby maintaining data confidentiality and integrity.

When the correct passphrase is entered, the \texttt{GPG} tool automatically detects the encryption algorithm (\texttt{AES256.CFB}) and completes the decryption process successfully. 
After decryption, the resulting plaintext backup file (\texttt{medibank\_backup.sql}) can be read and imported into the database. 
This demonstrates that the encryption and decryption mechanisms are functioning correctly and that the system's data backup security chain is fully operational.

\begin{figure}[htbp]
	\centering
	\includegraphics[width=0.45\textwidth]{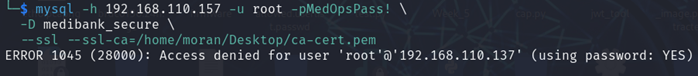}
	\caption{Disable Root Remote Login}
	\label{fig:drrl}
\end{figure}

Figure~\ref{fig:drrl} shows the configuration process used to disable remote login for the MySQL root user. By executing the commands 
\texttt{DELETE FROM mysql.user WHERE User='root' AND Host!='localhost';} 
followed by 
\texttt{FLUSH PRIVILEGES;}, 
all non-local root entries are removed. This measure ensures that administrative access is restricted to the local host, eliminating potential exploitation of privileged credentials over the network and enforcing the principle of least privilege.

\subsection{Network Controls: VPN and Jump Host}
\begin{figure}[htbp]
	\centering
	\includegraphics[width=0.45\textwidth]{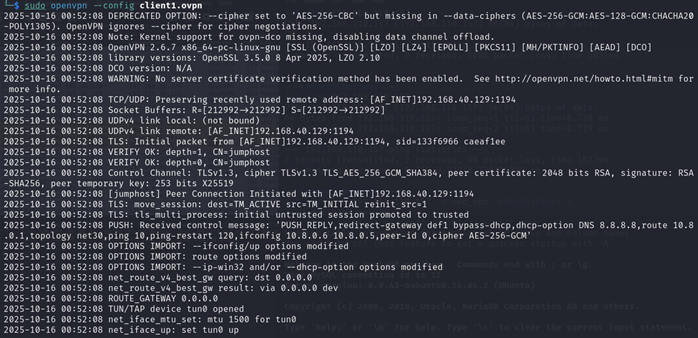}
	\caption{OpenVPN Connection Logs}
	\label{fig:ocl}
\end{figure}

Figure~\ref{fig:ocl} shows the client establishing a secure connection to the jump host using the OpenVPN protocol.
The console output indicates that TLS 1.3 was successfully negotiated with the cipher suite AES-256-GCM, and the VPN tunnel (tun0) was initialised without errors.
This confirms that the encrypted communication channel between the remote client (192.168.40.128) and the jump host (192.168.40.129) was successfully created, providing a trusted pathway for secure remote access to internal systems.

\begin{figure}[htbp]
	\centering
	\includegraphics[width=0.45\textwidth]{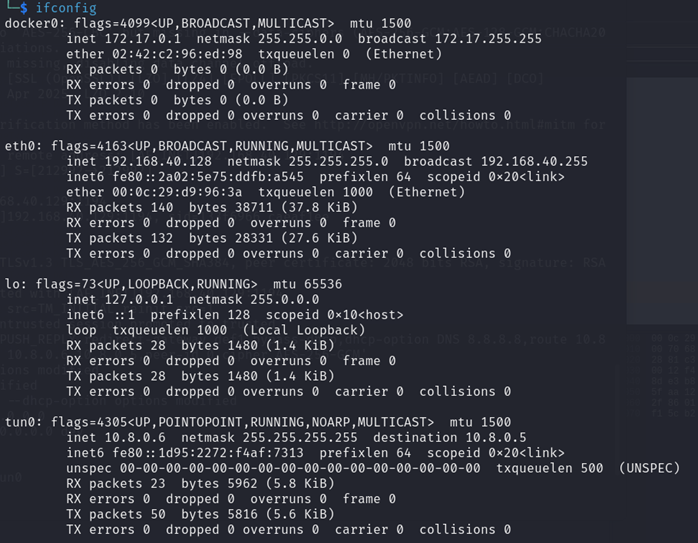}
	\caption{VPN Interface IP Configuration}
	\label{fig:viic}
\end{figure}

Figure~\ref{fig:viic} displays the network interface configuration (ifconfig) of the remote client after the OpenVPN session was established.
The newly created tun0 interface shows an assigned IP address in the 10.8.0.0/24 subnet, which corresponds to the VPN tunnel network.
This verifies that traffic from the client is being securely routed through the encrypted VPN channel to the internal network via the jump host, ensuring logical isolation from external networks.

\begin{figure}[htbp]
	\centering
	\includegraphics[width=0.45\textwidth]{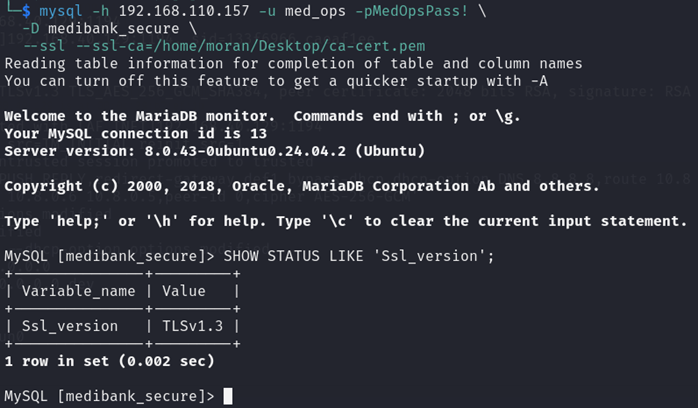}
	\caption{VPN Access To MySQL Successful}
	\label{fig:vatms}
\end{figure}

Figure~\ref{fig:vatms} demonstrates a successful connection from the remote client to the \texttt{medibank\_secure} MySQL database through the jump host using SSL/TLS encryption. The connection test executed with the account \texttt{med\_ops@\%} confirms proper authentication and encrypted data transmission under the TLSv1.3 protocol. This validates that the jump host effectively mediates secure, encrypted database access for authorised external users while enforcing network segmentation and credential controls.

\subsection{Python-Based Differential Privacy Implementation}
To demonstrate privacy-preserving analytics, we used the provided Python scripts (create\_dataset.py and differential\_privacy.py) to simulate a realistic deployment pipeline. The dataset generator creates different numbers of synthetic patient records (1000, 31000 and 131000) using the Faker library, which are stored in a SQLite database (medibank\_data.db) for testing. 

The differential privacy module applies six mechanisms shown in Table \ref{tab:dp-mechanisms}. 
\begin{table}[htbp]
	\caption{Differential Privacy Mechanisms and Applied Field}
	\label{tab:dp-mechanisms}
	\centering
	\begin{threeparttable}
		\renewcommand{\arraystretch}{1.15}
		\setlength{\tabcolsep}{4pt}
		\footnotesize
		\begin{tabular}{|>{\centering\arraybackslash}m{2.5cm}|
				>{\centering\arraybackslash}m{2.12cm}|
				>{\centering\arraybackslash}m{3.32cm}|}
			\hline
			\textbf{Mechanism} & \textbf{Applied Field} & \textbf{Purpose} \\ \hline
			Laplace Mechanism & DiagnosisCode & Adds noise to diagnosis counts to prevent re-identification \\ \hline
			Gaussian Mechanism & Age & Protects mean age computation from individual leakage \\ \hline
			Exponential Mechanism & TreatmentType & Randomizes treatment selection based on utility scores \\ \hline
			Randomised Response & AbortionFlag & Provides local privacy for sensitive binary attributes \\ \hline
			Histogram Mechanism & TreatmentType & Produces noisy distribution of treatment categories \\ \hline
			Sparse Vector Technique & Age & Efficiently answers threshold queries (e.g., age $>$ 65) \\ \hline
		\end{tabular}
	\end{threeparttable}
\end{table}

\begin{figure}[H]
	\centering
	\includegraphics[width=0.45\textwidth]{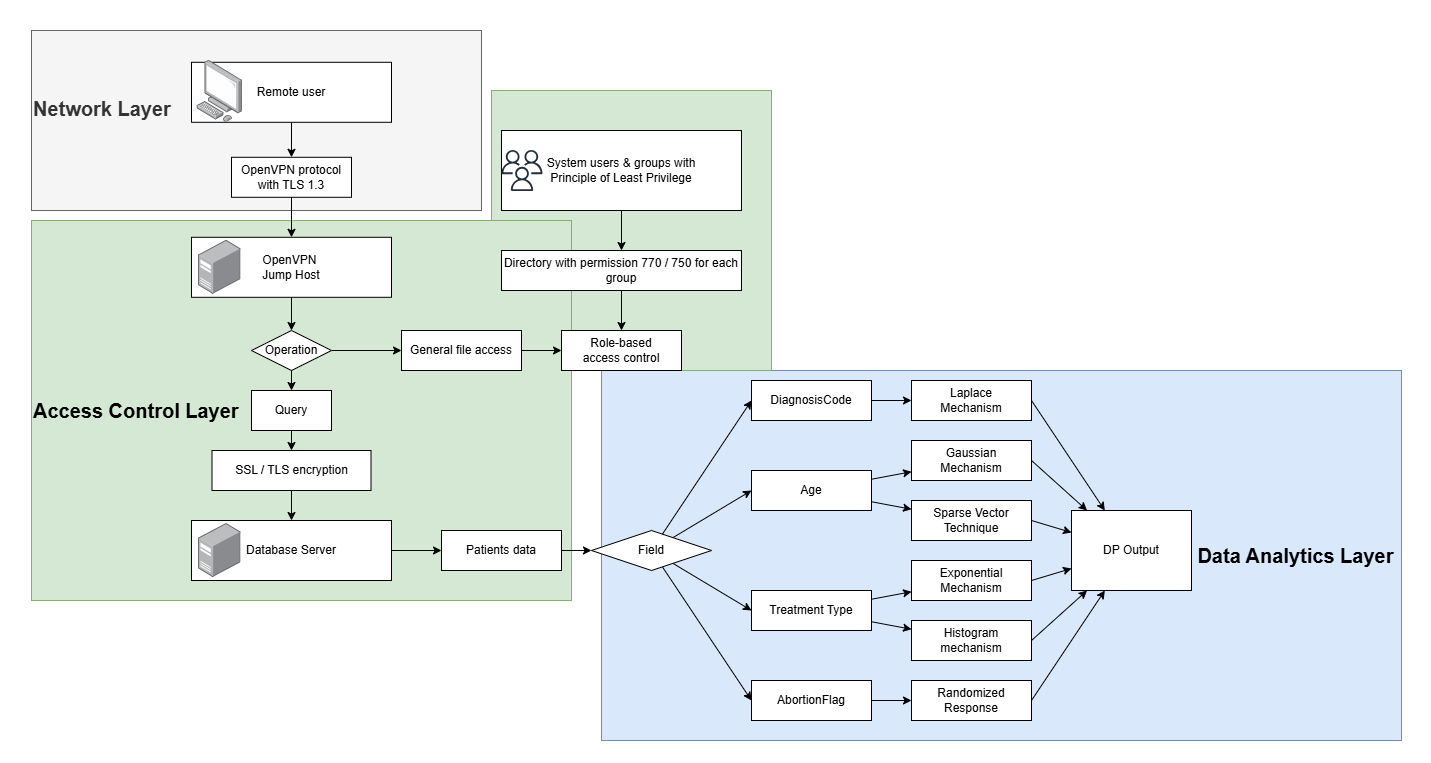}
	\caption{Field-to-Mechanism Mapping in Differential Privacy Workflow}
	\label{fig:fmmd}
\end{figure}

Figure~\ref{fig:fmmd} illustrates the mapping between medical data fields and their corresponding differential privacy mechanisms. Each sensitive attribute — such as DiagnosisCode, Age, TreatmentType, and AbortionFlag — is protected by a tailored privacy method (e.g., Laplace, Gaussian, Exponential, or Randomized Response). The workflow shows how raw patient data are processed through field-specific mechanisms to generate noise-injected outputs (Noised Data), ensuring both privacy protection and analytical utility.

The corresponding key Python code is shown in Listing~\ref{lst:laplace-gaussian},~\ref{lst:exp-rr} and~\ref{lst:hist-svt}.

\lstset{
	language=Python,
	basicstyle=\ttfamily\footnotesize,
	keywordstyle=\color{blue},
	commentstyle=\color{gray},
	stringstyle=\color{teal},
	breaklines=true,
	frame=single,
	numbers=none,
	showstringspaces=false,
	tabsize=4,
}

% --- Laplace Mechanism ---
\begin{lstlisting}[caption={Laplace and Gaussian Mechanisms}, label={lst:laplace-gaussian}]
	def laplace_mechanism(counts, sensitivity=1, epsilon=1.0):
	scale = sensitivity / epsilon
	noise = np.random.laplace(0, scale, size=len(counts))
	return counts + noise
	
	diagnosis_counts = df['DiagnosisCode'].value_counts()
	laplace_result = laplace_mechanism(diagnosis_counts, epsilon=1.0)
	
	def gaussian_mechanism(value, sensitivity=1, epsilon=1.0, delta=1e-5):
	sigma = np.sqrt(2 * np.log(1.25 / delta)) * sensitivity / epsilon
	noise = np.random.normal(0, sigma)
	return value + noise
	
	true_avg_age = df['Age'].mean()
	gaussian_result = gaussian_mechanism(true_avg_age, epsilon=1.0, delta=1e-5)
\end{lstlisting}

% --- Exponential Mechanism ---
\begin{lstlisting}[caption={Exponential and Randomized Response Mechanisms}, label={lst:exp-rr}]
	def exponential_mechanism(options, scores, epsilon=1.0):
	scores = np.array(scores)
	exp_scores = np.exp(epsilon * scores / (2 * max(scores)))
	probabilities = exp_scores / np.sum(exp_scores)
	selected = np.random.choice(options, p=probabilities)
	return selected, dict(zip(options, probabilities))
	
	treatment_counts = df['TreatmentType'].value_counts()
	treatment_options = treatment_counts.index.tolist()
	treatment_scores = treatment_counts.values.tolist()
	exp_result, exp_probs = exponential_mechanism(treatment_options, treatment_scores)
	
	def randomized_response(value, epsilon=1.0):
	p = np.exp(epsilon) / (1 + np.exp(epsilon))
	return value if np.random.rand() < p else 1 - value
	
	df['AbortionFlag'] = df['DiagnosisCode'].apply(lambda x: 1 if x == 'Z33.2' else 0)
	rr_flags = df['AbortionFlag'].apply(lambda x: randomized_response(x, epsilon=1.0))
	rr_mean = rr_flags.mean()
\end{lstlisting}

% --- Histogram Mechanism ---
\begin{lstlisting}[caption={Histogram and Sparse Vector Techniques}, label={lst:hist-svt}]
	def histogram_mechanism(counts, epsilon=1.0, clip_min=0):
	noisy_counts = {}
	for key, count in counts.items():
	noise = np.random.laplace(0, 1 / epsilon)
	noisy = max(count + noise, clip_min)
	noisy_counts[key] = noisy
	return noisy_counts
	
	histogram_result = histogram_mechanism(treatment_counts, epsilon=1.0)
	hist_entropy = safe_entropy(pd.Series(histogram_result))
	
	def sparse_vector_technique(values, threshold=65, epsilon=1.0, max_queries=10):
	results = []
	budget_used = 0
	for v in values:
	if budget_used >= max_queries:
	break
	noisy_v = v + np.random.laplace(0, 2 / epsilon)
	noisy_t = threshold + np.random.laplace(0, 4 / epsilon)
	results.append(noisy_v > noisy_t)
	budget_used += 1
	return results, budget_used
	
	svt_result, svt_budget = sparse_vector_technique(df['Age'].tolist())
\end{lstlisting}

This implementation establishes a reproducible pipeline for applying differential privacy mechanisms to sensitive medical attributes. 
The quantitative evaluation of this pipeline, including risk reduction and utility preservation, will be presented in Chapter~IV.

\subsection{Dataset Generation and Schema Design}

To simulate a realistic medical dataset for differential privacy testing, we implemented a local SQLite database using Python. The database schema, shown below, defines the \texttt{patients} table with nine fields, including demographic and medical attributes such as diagnosis and treatment type:

\begin{table}[htbp]
	\centering
	\begin{threeparttable}
		\renewcommand{\arraystretch}{1.15}
		\setlength{\tabcolsep}{4pt}
		\footnotesize
		\caption{Schema of the \texttt{patients} dataset}
		\label{tab:patients-schema}
		\begin{tabular}{|>{\centering\arraybackslash}m{1.9cm}|
				>{\centering\arraybackslash}m{1.8cm}|
				>{\centering\arraybackslash}m{2cm}|
				>{\centering\arraybackslash}m{2cm}|}
			\hline
			\textbf{Field Name} & \textbf{Data Type} & \textbf{Constraint} & \textbf{Description} \\
			\hline
			id & INTEGER & PRIMARY KEY, AUTOINCREMENT & Unique patient identifier \\
			\hline
			Name & TEXT & None & Patient name \\
			\hline
			Email & TEXT & None & Patient email address \\
			\hline
			DateOfBirth & TEXT & None & Date of birth (string) \\
			\hline
			MedicareNumber & TEXT & None & Medicare number \\
			\hline
			DiagnosisCode & TEXT & None & Diagnosis code \\
			\hline
			TreatmentType & TEXT & None & Treatment type \\
			\hline
			Address & TEXT & None & Residential address \\
			\hline
			Phone & TEXT & None & Contact phone number \\
			\hline
		\end{tabular}
	\end{threeparttable}
\end{table}

We use two controlled categorical fields \texttt{DiagnosisCode} and \texttt{TreatmentType} to ensure consistent data distribution across experiments. 
The diagnosis codes follow ICD-10 standards, while treatment types represent common medical procedures and interventions used in hospital systems, as shown in Listing~\ref{lst:init-diagnosis-treatment}.

\begin{lstlisting}[language=Python, caption={Initialization of diagnosis and treatment type arrays}, label={lst:init-diagnosis-treatment}]
	diagnosis_codes = ['F32.1', 'E11.9', 'Z33.2', ...]
	treatment_types = ['Pharmacotherapy', 'Radiotherapy', 'Chemotherapy', ...]
\end{lstlisting}

To populate the database, we employed the \texttt{Faker} library to generate synthetic yet realistic data records. Faker is a Python package for generating fake but semantically valid information, such as names, email addresses, dates of birth, and contact details, which makes it particularly suitable for privacy-preserving experiments. The key Python code is shown in Listing~\ref{lst:generate-medibank}.

\begin{lstlisting}[language=Python, caption={Function for Generating Synthetic Medibank Dataset}, label={lst:generate-medibank}]
	def generate_medibank_dataset(n=1000):
	records = []
	for _ in range(n):
	records.append({
		'Name': fake.name(),
		'Email': fake.email(),
		'DateOfBirth': fake.date_of_birth(minimum_age=18, maximum_age=90).strftime('%Y-%m-%d'),
		'MedicareNumber': generate_medicare_number(),
		'DiagnosisCode': random.choice(diagnosis_codes),
		'TreatmentType': random.choice(treatment_types),
		'Address': fake.address().replace('\n', ', '),
		'Phone': fake.phone_number()
	})
	return pd.DataFrame(records)
\end{lstlisting}

\begin{figure}[htbp]
	\centering
	\includegraphics[width=0.45\textwidth]{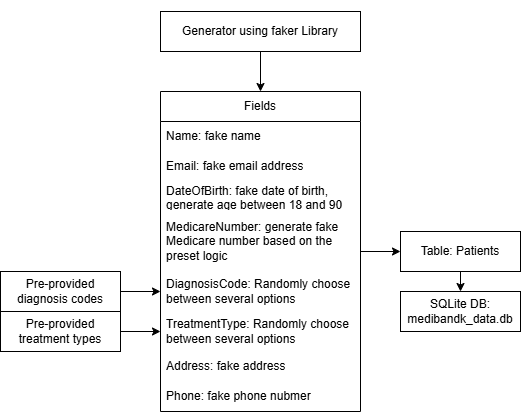}
	\caption{Workflow of Synthetic Medibank Dataset Generation using Faker Library}
	\label{fig:wosmd}
\end{figure}

Figure~\ref{fig:wosmd} illustrates the process of generating a synthetic Medibank dataset using the \texttt{Faker} library. Each record includes realistic demographic and medical attributes such as name, diagnosis, and treatment type. Controlled categorical fields (\texttt{DiagnosisCode} and \texttt{TreatmentType}) are pre-defined to ensure consistent distribution, while the final dataset is stored in an SQLite database (\texttt{medibank\_data.db}).

\section{Results and Evaluation}
\subsection{Differential Privacy Evaluation}
This section mainly assesses the effect of differential privacy technology. We use an evaluation script written in Python code to conduct a multi-faceted assessment of the effect of differential privacy. The following evaluation tables and figures are based on the operation and evaluation results of the differential privacy method on datasets of 1,000, 31,000, and 131,000. 

Table~\ref{tab:entropy-compare} compares the true entropy and noisy entropy of two key fields---\texttt{DiagnosisCode} and \texttt{TreatmentType}---for datasets of varying sizes. The results show that the entropy of \texttt{DiagnosisCode} remains nearly constant across different dataset sizes, whereas the entropy of \texttt{TreatmentType} decreases significantly after noise perturbation. This demonstrates that the noise mechanism has a more pronounced effect on categorical attributes, enhancing privacy while maintaining a relatively stable overall information distribution.

Table~\ref{tab:divergence-summary} presents the Kullback--Leibler (KL) and Jensen--Shannon (JS) divergence results for a hybrid Laplace and Exponential (Lap+Exp) mechanism applied to datasets of three different sizes. Both divergence metrics decrease consistently with increasing dataset size, from approximately $10^{-4}$ to $10^{-8}$. This trend indicates that larger datasets experience smaller overall distribution shifts after noise injection, achieving a more stable balance between privacy and utility.

Table~\ref{tab:reid-metrics} reports the results of re-identification attack evaluations on datasets of varying sizes. The key evaluation metrics include accuracy, precision, and recall. As dataset size increases, accuracy remains around 0.73, while precision exhibits a slight decrease. This indicates that although the model’s recognition capability remains relatively stable, the increased noise and anonymization reduce individual re-identifiability, thereby strengthening privacy protection.

Table~\ref{tab:mechanism-summary} provides a comparative overview of different differential privacy mechanisms (Laplace, Gaussian, Exponential, Randomised Response, Histogram, and Sparse Vector) evaluated under a privacy budget of $\varepsilon = 1$. The table includes utility loss, entropy change, and the query types best suited for each mechanism. Results show that the Laplace mechanism introduces minimal information loss in count queries, the Gaussian mechanism performs best for numerical queries, and the Histogram mechanism achieves a good trade-off between privacy and utility for statistical analyses. The Sparse Vector mechanism is particularly suitable for threshold queries. Overall, all mechanisms exhibit stable privacy performance as dataset size increases.

Table~\ref{tab:enh-mechanism-summary} compares the results of improved privacy-enhancing mechanisms across multiple algorithms (Laplace, Gaussian, Exponential, RR, Histogram, and Sparse) and dataset scales. The evaluation considers utility loss, entropy change, and privacy level. Experimental findings indicate that the enhanced mechanisms reduce utility loss while maintaining the same privacy budget. The Laplace and Gaussian mechanisms show the smallest entropy deviations, confirming that the improvements effectively mitigate information leakage. Moreover, these enhanced mechanisms demonstrate consistent scalability and robustness across different dataset sizes.

\begin{table}[htbp]
	\caption{Entropy Comparison}
	\label{tab:entropy-compare}
	\centering
	\begin{threeparttable}
		\renewcommand{\arraystretch}{1.1}
		\setlength{\tabcolsep}{4pt}
		\footnotesize
		\begin{tabular}{|>{\centering\arraybackslash}m{2.1cm}|
				>{\centering\arraybackslash}m{2.05cm}|
				>{\centering\arraybackslash}m{2.05cm}|
				>{\centering\arraybackslash}m{1.5cm}|}
			\hline
			\textbf{Field} & \textbf{True Entropy} & \textbf{Noisy Entropy} & \textbf{Dataset Size} \\ \hline
			DiagnosisCode & 4.38111264 & 4.380134325 & \multirow{2}{*}{1000} \\ \cline{1-3}
			TreatmentType & 4.69036867 & 3.497259978 & \\ \hline
			DiagnosisCode & 4.391838287 & 4.391833688 & \multirow{2}{*}{31000} \\ \cline{1-3}
			TreatmentType & 4.699843218 & 3.555122451 & \\ \hline
			DiagnosisCode & 4.392109341 & 4.392108768 & \multirow{2}{*}{131000} \\ \cline{1-3}
			TreatmentType & 4.700194665 & 3.569738249 & \\ \hline
		\end{tabular}
	\end{threeparttable}
\end{table}

\begin{table}[htbp]
	\caption{Divergence Summary}
	\label{tab:divergence-summary}
	\centering
	\begin{threeparttable}
		\renewcommand{\arraystretch}{1.1}
		\setlength{\tabcolsep}{4pt}
		\footnotesize
		\begin{tabular}{|>{\centering\arraybackslash}m{1.9cm}|
				>{\centering\arraybackslash}m{2.3cm}|
				>{\centering\arraybackslash}m{2.4cm}|
				>{\centering\arraybackslash}m{1.1cm}|}
            \hline
				\textbf{Mechanism} & \textbf{KL Divergence} & \textbf{JS Divergence} & \textbf{Dataset Size} \\ \hline
				Laplace and Exponential 
				& 0.000703426 
				& 0.000176109 
				& 1000 \\ \hline
				Laplace and Exponential 
				& 0.0000006899\newline60830072914 
				& 0.0000001724\newline82021423084 
				& 31000 \\ \hline
				Laplace and Exponential 
				& 0.0000000696\newline9996804998536 
				& 0.0000000174\newline25652516589637 
				& 131000 \\ \hline
		\end{tabular}
	\end{threeparttable}
\end{table}

\begin{table}[htbp]
	\caption{Reidentification Metrics}
	\label{tab:reid-metrics}
	\centering
	\begin{threeparttable}
		\renewcommand{\arraystretch}{1.1}
		\setlength{\tabcolsep}{4pt}
		\footnotesize
		\begin{tabular}{|>{\centering\arraybackslash}m{2.4cm}|
				>{\centering\arraybackslash}m{3.4cm}|
				>{\centering\arraybackslash}m{2.2cm}|}
			\hline
			\textbf{Metric} & \textbf{Value} & \textbf{Dataset Size} \\ \hline
			Accuracy & 0.738 & \multirow{3}{*}{1000} \\ \cline{1-2}
			Precision & 0.13286713286713286 & \\ \cline{1-2}
			Recall & 0.7307692307692307 & \\ \hline
			Accuracy & 0.7272258064516129 & \multirow{3}{*}{31000} \\ \cline{1-2}
			Precision & 0.12045032243961089 & \\ \cline{1-2}
			Recall & 0.7293183322303111 & \\ \hline
			Accuracy & 0.7308320610687022 & \multirow{3}{*}{131000} \\ \cline{1-2}
			Precision & 0.119138885159621948 & \\ \cline{1-2}
			Recall & 0.7264286857691692 & \\ \hline
		\end{tabular}
	\end{threeparttable}
\end{table}

\begin{table}[htbp]
	\caption{Mechanism Summary}
	\label{tab:mechanism-summary}
	\centering
	\begin{threeparttable}
		\renewcommand{\arraystretch}{1.2}
		\setlength{\tabcolsep}{4pt}
		\footnotesize
		\begin{tabular}{|>{\centering\arraybackslash}m{1.4cm}|
				>{\centering\arraybackslash}m{0.7cm}|
				>{\centering\arraybackslash}m{1.40cm}|
				>{\centering\arraybackslash}m{1.58cm}|
				>{\centering\arraybackslash}m{1.36cm}|
				>{\centering\arraybackslash}m{0.76cm}|}
			\hline
			\textbf{Mechanism} & \textbf{Privacy Level ($\varepsilon$)} & \textbf{Utility Loss} & \textbf{Entropy Change} & \textbf{Suitable For} & \textbf{Dataset Size} \\ \hline
			
			Laplace & 1 & 1.1723750\newline970886317 & 0.000978315 & Count Queries & \multirow{6}{*}{1000} \\ \cline{1-5}
			Gaussian & 1 & 1.3879380\newline842218652 & -- & Numeric Queries & \\ \cline{1-5}
			Exponential & 1 & Low & -- & Categorical Selection & \\ \cline{1-5}
			Randomized Response & 1 & 0.268 & -- & Binary Flags & \\ \cline{1-5}
			Histogram & 1 & 0.951869453 & 1.193108692 & Histograms & \\ \cline{1-5}
			Sparse Vector & 1 & Binary & -- & Threshold Queries & \\ \hline
			
			Laplace & 1 & 1.1441984\newline906362752 & 0.0000045990\newline9999502698 & Count Queries & \multirow{5}{*}{31000} \\ \cline{1-5}
			Gaussian & 1 & 4.0640324\newline78595699 & -- & Numeric Queries & \\ \cline{1-5}
			Exponential & 1 & Low & -- & Categorical Selection & \\ \cline{1-5}
			Randomized Response & 1 & 0.2674516\newline1290322583 & -- & Binary Flags & \\ \cline{1-5}
			Histogram & 1 & 1.1461431\newline185197875 & 1.144720767 & Histograms & \\ \hline
			
			Sparse Vector & 1 & Binary & 0.0000045990\newline9999502698 & Threshold Queries & \multirow{6}{*}{131000} \\ \cline{1-5}
			Laplace & 1 & 0.9245935\newline329410626 & 0.0000005730\newline000003367763 & Count Queries & \\ \cline{1-5}
			Gaussian & 1 & 3.6184728\newline290242774 & -- & Numeric Queries & \\ \cline{1-5}
			Exponential & 1 & Low & -- & Categorical Selection & \\ \cline{1-5}
			Randomized Response & 1 & 0.2702824\newline427480916 & -- & Binary Flags & \\ \cline{1-5}
			Histogram & 1 & 0.7939397\newline189997873 & 1.1304564162 & Histograms & \\ \hline
		\end{tabular}
	\end{threeparttable}
\end{table}

\begin{table}[htbp]
	\caption{Enhanced Mechanism Summary}
	\label{tab:enh-mechanism-summary}
	\centering
	\begin{threeparttable}
		\renewcommand{\arraystretch}{1.15}
		\setlength{\tabcolsep}{3pt}
		\footnotesize
		\begin{tabular}{|>{\centering\arraybackslash}m{1.6cm}|
				>{\centering\arraybackslash}m{2.2cm}|
				>{\centering\arraybackslash}m{2.2cm}|
				>{\centering\arraybackslash}m{0.8cm}|
				>{\centering\arraybackslash}m{1.0cm}|}
			\hline
			\textbf{Mechanism} & \textbf{Utility Loss} & \textbf{Entropy Change} & \textbf{Privacy Level} & \textbf{Dataset Size} \\ \hline
			
			Laplace & 0.85049522\newline42686175 & -0.001152546\newline00862702 & 1 & \multirow{6}{*}{1000} \\ \cline{1-4}
			Gaussian & 0.5278533\newline639775702 & 0 & 1 & \\ \cline{1-4}
			Exponential & 0.1 & 0 & 1 & \\ \cline{1-4}
			RR & 0.25 & 0 & 1 & \\ \cline{1-4}
			Histogram & 0.6 & 4.6903686\newline6969128 & 1 & \\ \cline{1-4}
			Sparse & 0.1 & 0 & 1 & \\ \hline
			
			Laplace & 1.5994968\newline859359429 & -0.0000070358\newline44953762194 & 1 & \multirow{6}{*}{31000} \\ \cline{1-4}
			Gaussian & 0.61507410\newline06547378 & 0 & 1 & \\ \cline{1-4}
			Exponential & 0.1 & 0 & 1 & \\ \cline{1-4}
			RR & 0.25 & 0 & 1 & \\ \cline{1-4}
			Histogram & 0.6 & 4.6998432\newline17542753 & 1 & \\ \cline{1-4}
			Sparse & 0.1 & 0 & 1 & \\ \hline
			
			Laplace & 1.21704853\newline76394762 & -0.0000034560\newline953752205137 & 1 & \multirow{6}{*}{131000} \\ \cline{1-4}
			Gaussian & 5.5410819\newline68129745 & 0 & 1 & \\ \cline{1-4}
			Exponential & 0.1 & 0 & 1 & \\ \cline{1-4}
			RR & 0.25 & 0 & 1 & \\ \cline{1-4}
			Histogram & 0.6 & 4.7001946\newline65224181 & 1 & \\ \cline{1-4}
			Sparse & 0.1 & 0 & 1 & \\ \hline
		\end{tabular}
	\end{threeparttable}
\end{table}

\begin{figure}[htbp]
	\centering
	\includegraphics[width=0.45\textwidth]{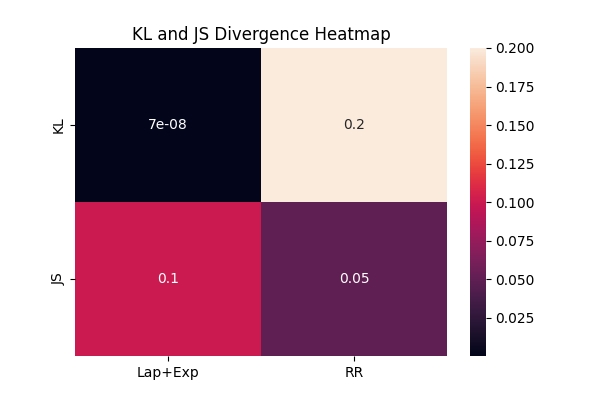}
	\caption{Divergence Heatmap}
	\label{fig:dh}
\end{figure}

Figure~\ref{fig:dh} compares the distribution-preserving capabilities of two privacy-preserving mechanisms, using the KL and JS divergences to measure the difference in distribution after data perturbation. The horizontal axis represents the mechanism type: a combination of Laplace and Exponential mechanisms (Lap+Exp) and a Randomised Response (RR) mechanism; the vertical axis represents the corresponding divergence type. Colors reflect numerical values, with lighter colors indicating greater distribution shifts. The KL divergence of the Lap+Exp mechanism is approximately $4.7 \times 10^{-8}$ and the JS divergence is $1.2 \times 10^{-8}$, confirming that Lap+Exp better maintains overall distribution consistency while perturbing the data. This result is consistent with the experimental analysis of differential privacy in this paper's medical data leakage prevention, demonstrating that noise injection-based mechanisms offer advantages in balancing data availability and privacy, and are particularly applicable to large-scale health data exposures like Medibank.

\begin{figure}[htbp]
	\centering
	\includegraphics[width=0.45\textwidth]{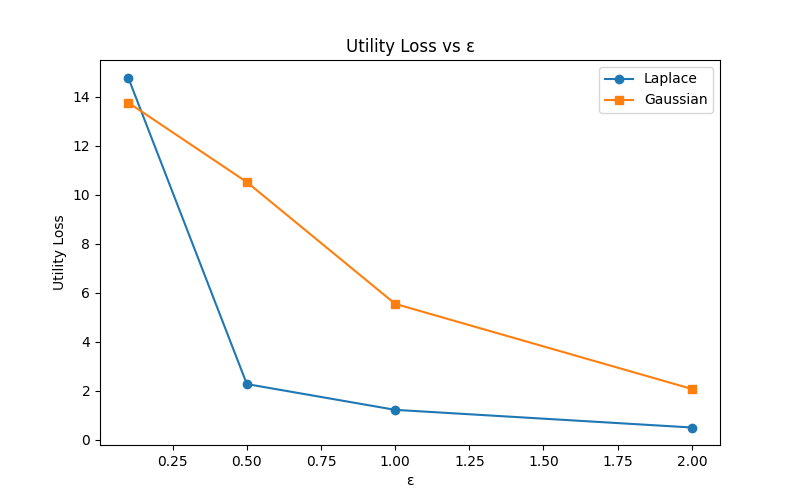}
	\caption{Epsilon vs Loss}
	\label{fig:evl}
\end{figure}

Figure~\ref{fig:evl} shows the changing trends in data utility loss for the Laplace and Gaussian differential privacy mechanisms at different privacy strengths ($\varepsilon$ values). The horizontal axis represents the privacy budget $\varepsilon$, and the vertical axis represents the corresponding utility loss. As the $\varepsilon$ value increases (i.e., the degree of privacy protection decreases), the utility loss of both mechanisms decreases significantly. The utility loss of the Laplace mechanism decreases more rapidly, from approximately $0.9$ at $\varepsilon=1.0$ to near zero at higher $\varepsilon$ values. The Gaussian mechanism, on the other hand, has a higher overall utility loss (around $3.3$ at $\varepsilon=1.0$), and the downward trend is more gradual. This demonstrates that, for the same privacy budget, the Laplace mechanism can provide stronger privacy protection while maintaining higher data availability. This result confirms the effectiveness of the differential privacy optimization scheme proposed in this paper in balancing privacy and utility, providing feasible parameter guidance for medical data scenarios (such as the Medibank case).

\begin{figure}[H]
	\centering
	\includegraphics[width=0.45\textwidth]{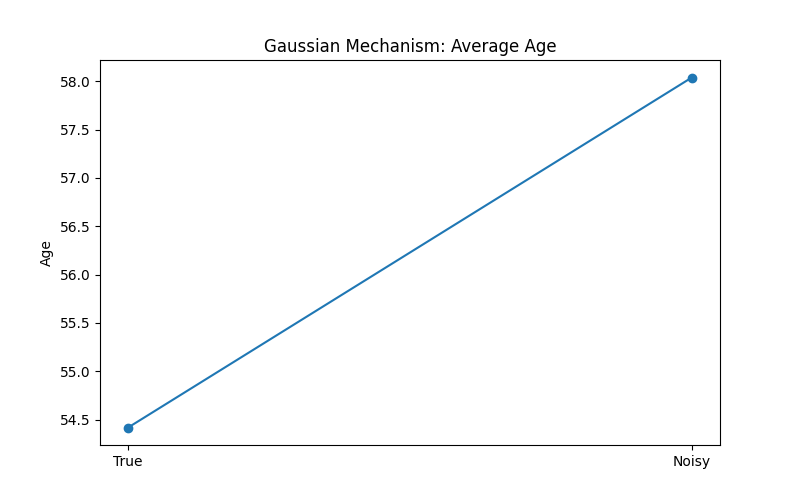}
	\caption{Gaussian Age}
	\label{fig:ga}
\end{figure}

Figure~\ref{fig:ga} shows the change in the average age in the dataset after applying the Gaussian mechanism. The horizontal axis represents the data type (true value and noisy value), and the vertical axis represents the corresponding average age. As can be seen, the average age in the original data was approximately 54.4 years old, which increased to approximately 58.0 years old after the addition of Gaussian noise.

This change reflects the typical characteristics of the Gaussian mechanism in privacy protection: by superimposing random noise that conforms to a normal distribution on the original data, individual information is obscured, preventing the accurate reconstruction of sensitive statistics. Although the introduction of noise causes a slight shift in the mean, the overall data trend remains consistent. This demonstrates that in medical data scenarios, the Gaussian mechanism can achieve statistical privacy protection within an acceptable error range, allowing the data to be used for aggregate analysis without revealing individual characteristics.

\begin{figure}[htbp]
	\centering
	\includegraphics[width=0.5\textwidth]{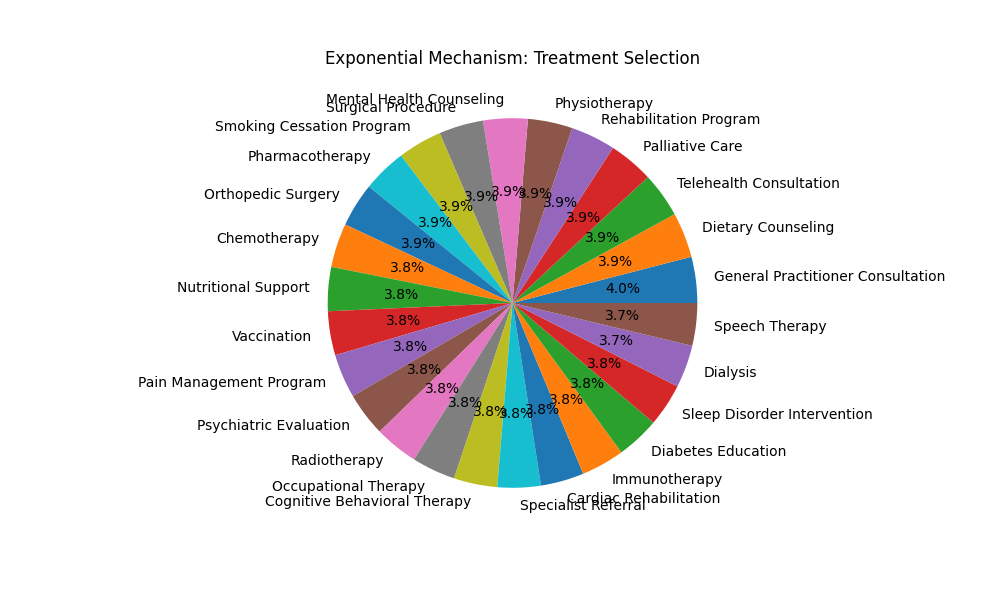}
	\caption{Exponential Treatment}
	\label{fig:et}
\end{figure}

Figure~\ref{fig:et} shows the distribution of selection probabilities for different treatment types after applying differential privacy using the Exponential Mechanism. The figure lists several typical medical services, such as general practitioner consultation, radiotherapy, mental health counseling, and vaccination, with each accounting for approximately 3.7\% to 4.0\% of the pie chart.

This balanced distribution indicates that, after the perturbation process using the Exponential Mechanism, the selection probabilities of each treatment option are randomised to a near-uniform level, effectively preventing over-exposure of sensitive treatment records. Compared to the likely high-frequency treatment types in the original dataset, the Exponential Mechanism, through exponentially weighted sampling of the utility function, ensures that each candidate option has a similar probability of selection. This result demonstrates the effectiveness of differential privacy in controlling sensitivity bias and improving individual privacy protection in the release of medical data, while maintaining the overall rationality of statistical analysis.

\begin{figure}[htbp]
	\centering
	\includegraphics[width=0.45\textwidth]{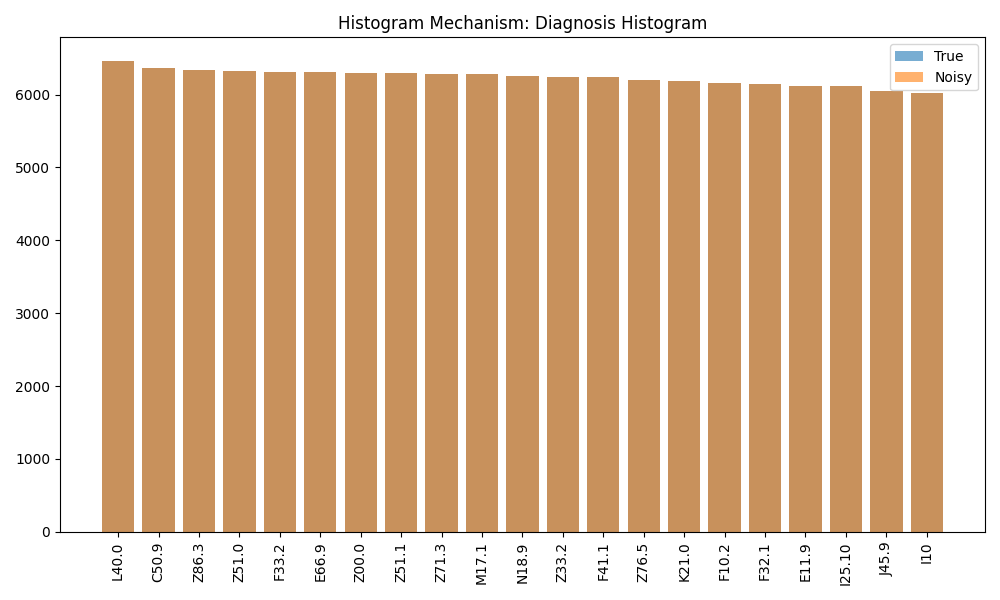}
	\caption{Histogram Diagnosis}
	\label{fig:hd}
\end{figure}

Figure~\ref{fig:hd} shows the frequency distribution of different diagnosis codes in real data and noisy data after processing using the histogram mechanism. The horizontal axis represents the diagnosis code (such as L40.0, C50.9, Z86.3, etc.), and the vertical axis represents the number of records corresponding to each code. The blue bars in the figure represent the real data (True), and the orange bars represent the data after adding noise (Noisy).

The results show that the heights of the real and noisy values are almost the same, with only very slight differences in individual categories. This indicates that the histogram mechanism effectively preserves the original data structure at the global distribution level while reducing the identifiability of individual records through random perturbations.

This figure shows the change in counts for different diagnosis codes after differential privacy processing using the Laplace mechanism. The horizontal axis represents the diagnosis code (e.g., L40.0, C50.9, Z86.3), and the vertical axis represents the number of records corresponding to each code. The blue bars in the figure represent true data, and the orange bars represent noisy data.

\begin{figure}[htbp]
	\centering
	\includegraphics[width=0.45\textwidth]{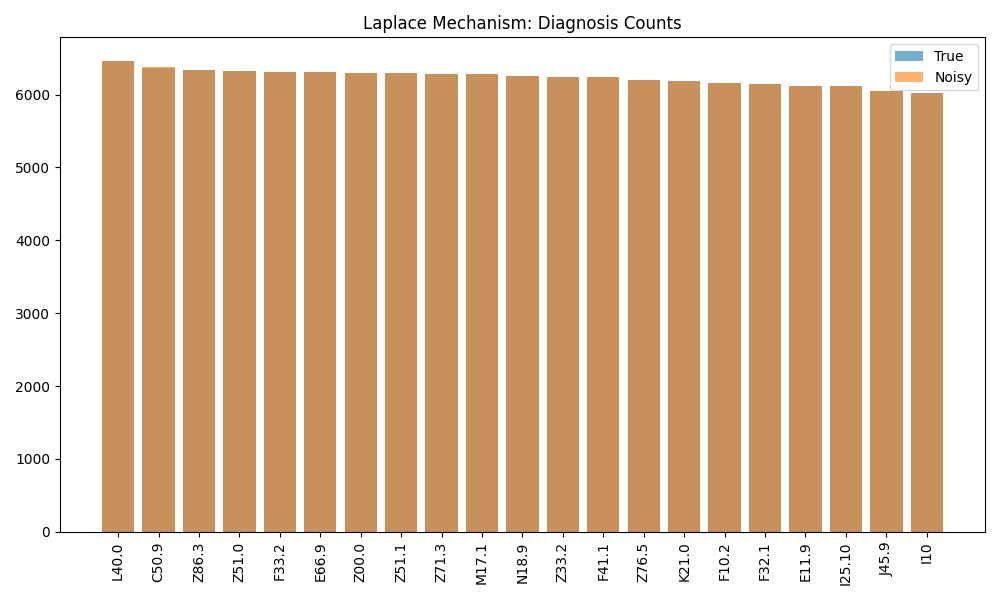}
	\caption{Laplace Diagnosis}
	\label{fig:ld}
\end{figure}

As can be seen from Figure~\ref{fig:ld}, the true and noisy values are almost entirely consistent, demonstrating that the Laplace mechanism maintains the overall statistical distribution structure well after adding noise, with only minor deviations in the values. This mechanism effectively hides individual contributions by superimposing random noise that conforms to a Laplace distribution on each count value, thereby preventing sensitive diagnosis records from being inferred or restored. This result demonstrates that the Laplace mechanism offers a high balance between data fidelity and privacy protection in medical data scenarios, ensuring the accuracy of statistical analysis while satisfying differential privacy constraints.

\begin{figure}[htbp]
	\centering
	\includegraphics[width=0.45\textwidth]{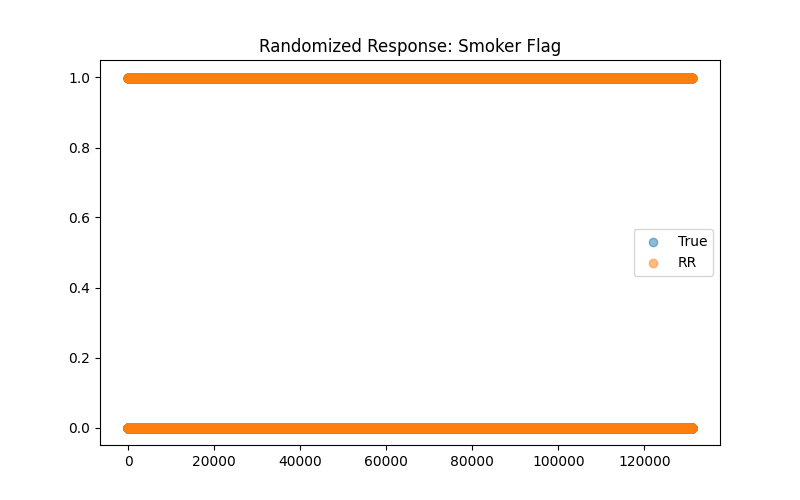}
	\caption{Randomized Response}
	\label{fig:rr}
\end{figure}

Figure~\ref{fig:rr} illustrates the effectiveness of the RR mechanism for processing the binary sensitive attribute ``Smoker Flag.'' The horizontal axis represents the sample size (approximately 131,000 records), and the vertical axis represents the binary status (0 for non-smoker, 1 for smoker). The blue dots in the figure represent the true data (True), while the orange dots represent the data after RR processing.

While the distribution after RR matches the overall shape of the true data, random flipping occurs at the local sample level—some samples labelled initially ``1'' (smoker) are perturbed to ``0,'' and vice versa. This randomisation process is the core of the RR mechanism. By introducing controlled probabilistic noise at the individual level, it prevents attackers from determining the true attributes of individual users while still ensuring the accuracy of the overall statistical proportions.

\begin{figure}[htbp]
	\centering
	\includegraphics[width=0.45\textwidth]{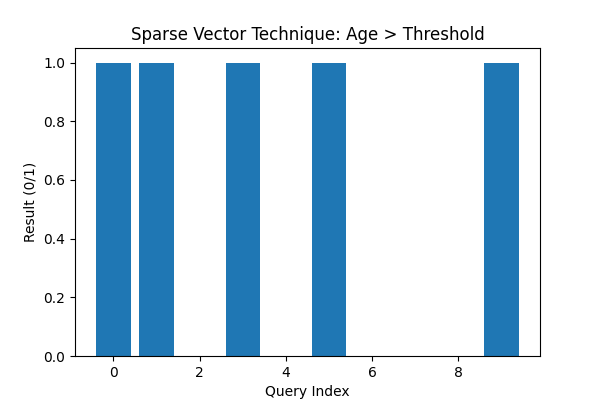}
	\caption{Sparse Vector}
	\label{fig:sv}
\end{figure}

Figure~\ref{fig:sv} illustrates the performance of the Sparse Vector Technique (SVT) when executing differentially private queries, specifically applying it to the task of determining whether age exceeds a threshold (Age $>$ Threshold).

The horizontal axis represents the query index, and the vertical axis represents the binary result (0 or 1), where 1 indicates that the query result exceeds the threshold, and 0 indicates that it does not. The figure shows several query operations, and we can observe that the majority of queries result in 1, while only a few fail to meet the criteria.

This result distribution demonstrates the characteristics of the Sparse Vector Technique:
\begin{itemize}
	\item SVT introduces controlled noise into each comparison, allowing the system to return true only when a small number of ``significant'' queries meet the condition, while suppressing or randomizing non-significant results.
	\item This mechanism manages the overall privacy budget while handling multiple rounds of queries, preventing privacy leaks caused by frequent access to sensitive data.
\end{itemize}

\begin{figure}[htbp]
	\centering
	\includegraphics[width=0.45\textwidth]{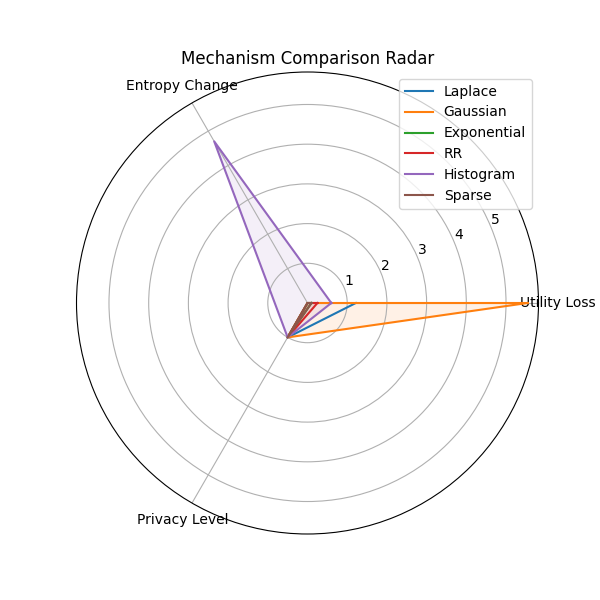}
	\caption{Radar Mechanism}
	\label{fig:rm}
\end{figure}

Figure~\ref{fig:rm} compares six differential privacy mechanisms across three dimensions: utility loss, entropy change, and privacy level, as visualised using a radar chart. The figure plots the performance profiles of the six mechanisms: Laplace, Gaussian, Exponential, Randomized Response (RR), Histogram, and Sparse Vector. 

The results reveal significant differences among these mechanisms in the three metrics. 
\begin{itemize}
	\item The Gaussian mechanism exhibits the highest utility loss, indicating that it introduces substantial data distortion under high privacy constraints.
	\item The Histogram mechanism shows the greatest entropy change, suggesting that it more strongly affects the distribution shape after data perturbation.
	\item The Laplace, Exponential, RR, and Sparse mechanisms, in contrast, display more compact profiles, reflecting a balanced trade-off between privacy protection and data usability.
\end{itemize}

Overall, the radar chart highlights the comparative performance of different differential privacy mechanisms for medical data protection. The Laplace mechanism is particularly suitable for count queries that require high precision, whereas the Gaussian mechanism is more appropriate for numerical analysis scenarios. Although the Histogram mechanism introduces higher distributional variation, it remains effective for statistical publishing tasks. This visual analysis provides practical guidance for selecting optimal privacy-preserving strategies in medical data security frameworks, such as those designed to prevent data breaches in Medibank healthcare systems.

\begin{figure}[htbp]
	\centering
	\includegraphics[width=0.45\textwidth]{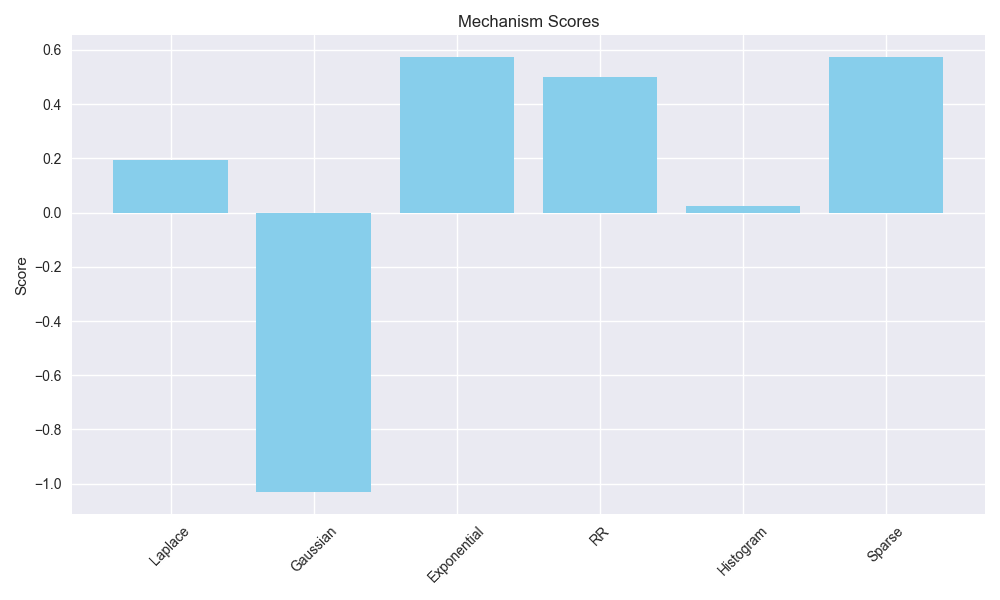}
	\caption{Score Bar}
	\label{fig:sb}
\end{figure}

Figure~\ref{fig:sb} shows the scores of different differential privacy mechanisms under a comprehensive performance evaluation. The horizontal axis represents the mechanism names, including Laplace, Gaussian, Exponential, RR, Histogram, and Sparse; the vertical axis represents the overall score, which measures each mechanism's balance between privacy protection and data availability.

As can be seen from the figure:
\begin{itemize}
	\item Exponential, Sparse, and RR mechanisms achieve the highest scores, approximately 0.5 to 0.6, indicating that they offer a good balance between privacy and utility under this evaluation framework and are particularly suitable for tasks involving categorical data or binary labels.
	\item The Laplace mechanism scores slightly lower (approximately 0.2), indicating a moderate trade-off between noise intensity and data availability, maintaining good versatility.
	\item The Histogram mechanism scores close to zero, indicating potential utility loss when applied to high-dimensional data.
	\item The Gaussian mechanism has a negative score (approximately $-1.0$), making it the worst-performing mechanism. This suggests that it introduces excessive noise under the experimental conditions, leading to data bias and reduced practicality.
\end{itemize}

\begin{figure}[htbp]
	\centering
	\includegraphics[width=0.45\textwidth]{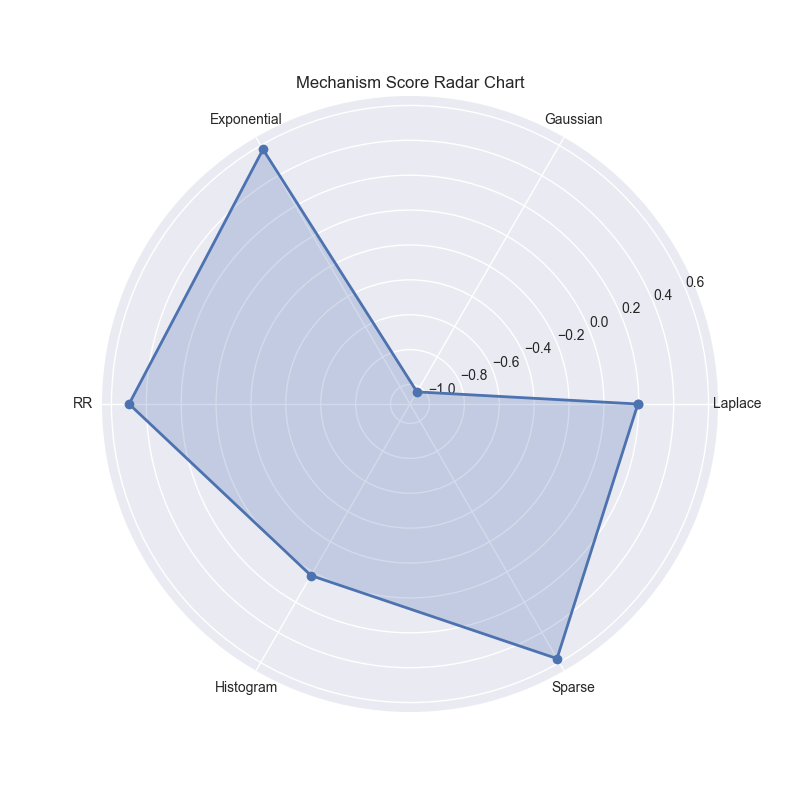}
	\caption{Score Radar}
	\label{fig:sr}
\end{figure}

Figure~\ref{fig:sr} is a differential privacy mechanism score radar chart, which visualises the relative performance of different privacy-preserving mechanisms in a comprehensive performance evaluation. The axes of the radar chart represent different differential privacy mechanisms, including Laplace, Gaussian, Exponential, RR, Histogram, and Sparse, with the radius representing their corresponding comprehensive scores (ranging from approximately $-1.0$ to $0.6$).

Figure~\ref{fig:sr} illustrates the following characteristics:
\begin{itemize}
	\item The Exponential mechanism has the highest score among all mechanisms, close to $0.6$, indicating that it performs best in balancing privacy and utility.
	\item The RR and Sparse Vector techniques follow closely behind, also achieving high scores, demonstrating that they strike a good balance between data perturbation and interpretability.
	\item The Laplace mechanism has a moderate score (approximately $0.2$–$0.3$), reflecting its stability in numerical data scenarios, but it performs slightly worse compared to the Exponential mechanism.
	\item The Histogram mechanism scores low, suggesting that it may introduce significant noise in high-dimensional or multi-class data scenarios.
	\item The Gaussian mechanism scores the lowest (close to $-1.0$), demonstrating significantly poor performance, as it suffers from strong noise interference that severely reduces data utility.
\end{itemize}

\subsection{Database Security Validation}

\begin{figure}[htbp]
	\centering
	\includegraphics[width=0.45\textwidth]{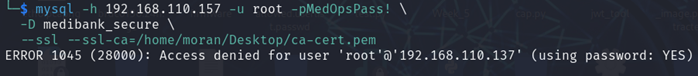}
	\caption{Remote Database Connection Failure Example}
	\label{fig:rdcfe}
\end{figure}

Figure~\ref{fig:rdcfe} shows the result of an attempted remote login to the MySQL database using the root account, which was denied with an ``Access denied for user 'root'@'192.168.110.157' '' error. This result indicates that remote access for the root user has been successfully disabled in accordance with security best practices. Restricting root access to local connections helps prevent unauthorised remote control of the database, reducing the attack surface and protecting sensitive medical data stored in the Medibank system.

\begin{figure}[htbp]
	\centering
	\includegraphics[width=0.45\textwidth]{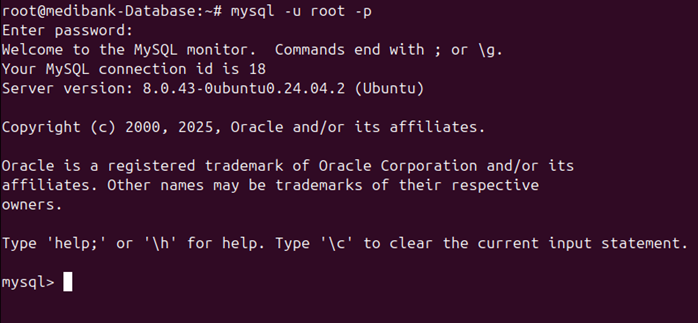}
	\caption{Example of Successful Local Database Login}
	\label{fig:eosldl}
\end{figure}

Figure~\ref{fig:eosldl} shows the result of a successful local login to the MySQL database from the same host using the root account. This demonstrates that while remote root access is blocked, local administrative access remains functional, allowing authorised system administrators to manage the database securely from within the server environment. This configuration achieves a secure balance between administrative usability and remote access control, reinforcing the system’s defence against external intrusion.

\subsection{Ethical Hacking Penetration Testing}
After we have completed the deployment and security configuration of the system and software, we will conduct ethical hacking tests to determine whether they can successfully prevent attacks. We will use several tools, such as nmap, sqlmap, and Wireshark, which are already integrated in the Kali system, to conduct a series of penetration tests on the security of the MySQL database, where Ubuntu 24.04 is located, to see if the security measures we implement can pass the penetration test when exporting or operating on the contents of the database.

\begin{figure}[htbp]
	\centering
	\includegraphics[width=0.45\textwidth]{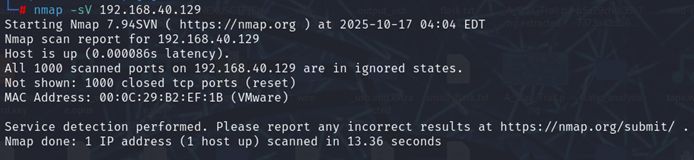}
	\caption{Nmap Scan Result of Jump Host}
	\label{fig:nsr}
\end{figure}

First, we use nmap to scan the Host. Since there is a jump host in the middle, we can only scan the information of the jump host, and the scan results are shown in Figure~\ref{fig:nsr}. We cannot obtain the information from the MySQL server. Since connecting to the jump host requires the OpenVPN client private key credentials, and it’s the only way we can access the MySQL server, we cannot connect to or scan the MySQL server directly.

\begin{figure}[htbp]
	\centering
	\includegraphics[width=0.45\textwidth]{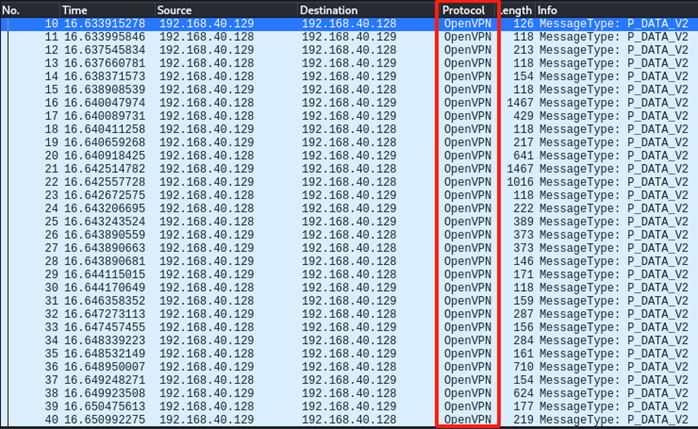}
	\caption{OpenVPN Encrypted Tunnel Traffic Between Jump Host and Client}
	\label{fig:oett}
\end{figure}

Next, we test the scenario of a man-in-the-middle attack. Suppose someone connects to the MySQL host through the jump host to log in to the database and perform some operations. At this time, we act as a man-in-the-middle and use Wireshark to capture the communication data packets to see if we can capture some useful information. The packet capture result is shown in Figure~\ref{fig:oett}. The results show that although we have captured the data packets of database operations, the real request contents within them cannot be seen because the VPN protocol has encrypted them.

\begin{figure}[htbp]
	\centering
	\includegraphics[width=0.45\textwidth]{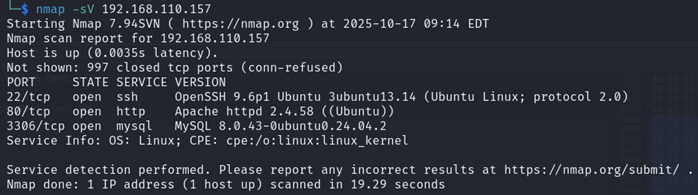}
	\caption{Nmap Service Scan of Database Server}
	\label{fig:nss}
\end{figure}

\begin{figure}[htbp]
	\centering
	\includegraphics[width=0.45\textwidth]{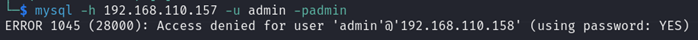}
	\caption{Remote Login Attempt with Denied Access for Admin User}
	\label{fig:rla}
\end{figure}

Extreme Case 1: Suppose in the extreme case, the authorised VPN credentials are stolen (the same situation as the Medibank 2022 data breach). We can use the VPN credentials to connect to the jump host and scan it again with nmap. The result is as shown in Figure~\ref{fig:nss}. At this time, the MySQL server's information can be scanned, but a connection cannot be initiated because the MySQL database has enabled forced TLS/SSL login. Since the attacker doesn’t have account information or passwords, the attacker can only try brute force cracking. However, because forced TLS/SSL login is enabled, brute-force cracking cannot be performed as the server will directly reject any requests without a private key. The result is as shown in Figure~\ref{fig:rla}. In this way, in remote scenarios, our software and system security is very effective. 

\begin{figure}[htbp]
	\centering
	\includegraphics[width=0.45\textwidth]{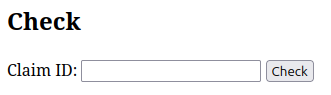}
	\caption{SQL Injection Test Interface}
	\label{fig:sqliti}
\end{figure}

\begin{figure}[htbp]
	\centering
	\includegraphics[width=0.45\textwidth]{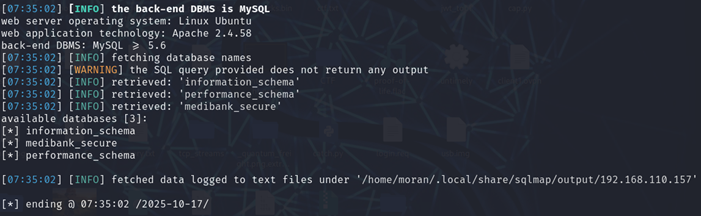}
	\caption{SQLMap Database Enumeration Result}
	\label{fig:sde}
\end{figure}

\begin{figure}[htbp]
	\centering
	\includegraphics[width=0.45\textwidth]{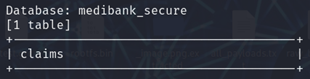}
	\caption{Retrieved Table List in \texttt{medibank\_secure} Database}
	\label{fig:rtl}
\end{figure}

\begin{figure}[htbp]
	\centering
	\includegraphics[width=0.45\textwidth]{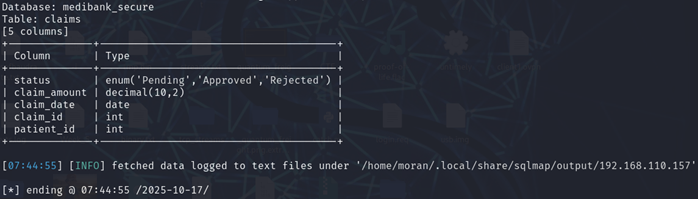}
	\caption{Table Schema Enumeration for \texttt{claims} Table}
	\label{fig:tse}
\end{figure}

\begin{figure}[htbp]
	\centering
	\includegraphics[width=0.45\textwidth]{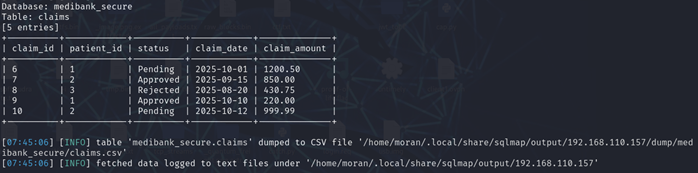}
	\caption{Data Extraction from \texttt{claims} Table}
	\label{fig:de}
\end{figure}

\begin{figure}[htbp]
	\centering
	\includegraphics[width=0.45\textwidth]{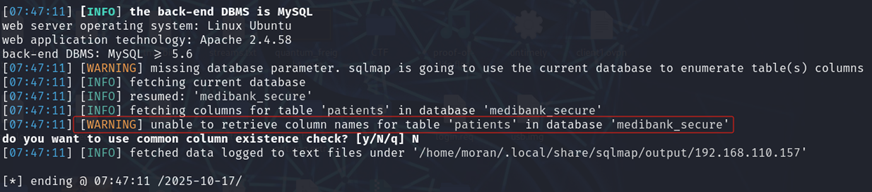}
	\caption{Failed Enumeration Attempt for \texttt{patients} Table}
	\label{fig:fea}
\end{figure}

Extreme Case 2: Injectable Website.
Next, let's assume another extreme case. We deploy the web application service on MySQL's native machine. We enable the Apache + PHP service on the MySQL server and deliberately place a PHP page with an SQL injection vulnerability for testing. To be more realistic, we store the \texttt{claims\_team} database user in PHP, as this user is more likely to have external web query purposes, such as querying the company's claims status. The attacker can directly access the web services without connecting to the jump host. The SQL injection page is shown in Figure~\ref{fig:sqliti}. Next, we used \texttt{sqlmap} to scan the injection points on the page and attempted SQL injection to test if we could operate or dump the database. The result is shown in Figure~\ref{fig:sde}. The results show that there is an SQL vulnerability, allowing access to all the databases in MySQL. Next, we attempt to view the tables in \texttt{medibank\_secure} database, and the result is shown in Figure~\ref{fig:rtl}. From the results, it can be seen that since the PHP file only contains the account and password of \texttt{claims\_team}, it only has the permission to view the \texttt{claims} table and cannot see the names of other tables in the database. We can view the structure of the \texttt{claims} table and export its data, as shown in Figure~\ref{fig:tse} and ~\ref{fig:de}. However, even though we know that the \texttt{patients} table exists, attempting to view its structure directly fails, as shown in Figure~\ref{fig:fea}. This indicates that our permission division and data segmentation are effective. Even without IDS/IPS and existing vulnerabilities, attackers still can’t view or export data beyond their permissions.

SQL injectable website shown in Figure~\ref{fig:sqliti}.

\begin{figure}[htbp]
	\centering
	\includegraphics[width=0.45\textwidth]{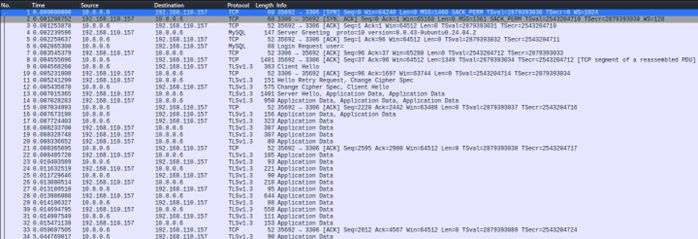}
	\caption{Captured TLS 1.3 Encrypted Communication Between Client and Database Server}
	\label{fig:ctec}
\end{figure}

\begin{figure}[htbp]
	\centering
	\includegraphics[width=0.45\textwidth]{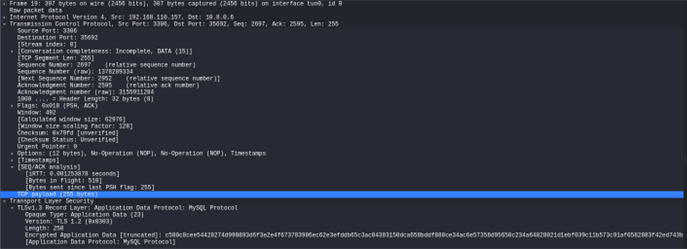}
	\caption{TLS Layer Inspection Showing Encrypted MySQL Application Data Payload}
	\label{fig:tlise}
\end{figure}

Extreme Case 3: We continue with an extreme test. It is still assumed that the attacker has stolen the VPN login credentials and is capturing the database operation information in the VPN channel. The result of using Wireshark to capture data packets is shown in Figure~\ref{fig:ctec}, and a detailed view of one of the packets is shown in Figure~\ref{fig:tlise}. It can be seen that TLS/SSL encrypts the data packets, yet no plaintext information is visible. Even under such extreme conditions, the attacker still cannot obtain any plaintext information or database information. Unless the private key file connected to the database is stolen, attackers may not be able to attempt a brute force attack on the MySQL database username and password. However, in our deployment plan, there is also MFA verification. Therefore, even if the private key and account password are known, MFA verification is still required to successfully log in to the database, dramatically increasing the difficulty of attacking it. Our solution effectively protects the database.

\subsection{Improvement Measures: Mean Age Differential Privacy Optimization}
In the original experiment, the use of the Gaussian Mechanism for mean age perturbation resulted in a significant mean shift (from 54.4 to approximately 58.0 years).  
Such deviation may cause distortion in population-level medical statistics, potentially compromising the reliability of downstream models and risk assessment.  
Therefore, this section proposes three improvement strategies designed to minimize the noise-induced bias in mean statistics while maintaining the privacy budget constraint.

\subsubsection{Standard Laplace Mechanism for Mean}
\textbf{Motivation:} Gaussian noise may cause bias in datasets with small sample size or low variance, whereas the Laplace mechanism under $\varepsilon$-DP provides a tighter privacy guarantee.  
By defining the sensitivity as $\Delta f = \frac{\text{max} - \text{min}}{n}$ and adding Laplace noise accordingly, a more stable mean estimate can be obtained. The key Python code is shown in Listing~\ref{lst:laplace-avg-age-global}.

\begin{lstlisting}[language=Python, caption={Laplace Mechanism for Noisy Average Age Computation}, label={lst:laplace-avg-age-global}]
	epsilon = 1.0
	true_avg_age = data['Age'].mean()
	sensitivity = 100 / len(data)
	laplace_noise = np.random.laplace(0, sensitivity / epsilon)
	noisy_avg_age = true_avg_age + laplace_noise
\end{lstlisting}

As shown in Listing~\ref{lst:laplace-avg-age-global}, this mechanism injects noise based on \textit{global sensitivity}, making it suitable for datasets with relatively uniform distributions and insignificant outliers.

\subsubsection{Bounded Laplace Mechanism (Clipped Mean Age)}
\textbf{Motivation:}  
In medical datasets, extreme values (e.g., very young or very old patients) may inflate global sensitivity and lead to excessive noise.  
By clipping the age range (e.g., between 18 and 90), the sensitivity can be significantly reduced, producing more stable mean estimates. The key Python code is shown in Listing~\ref{lst:laplace-avg-age-clipped}.

\begin{lstlisting}[language=Python, caption={Clipped Laplace Mechanism for Average Age}, label={lst:laplace-avg-age-clipped}]
	clipped_age = data['Age'].clip(lower=18, upper=90)
	true_avg_clipped = clipped_age.mean()
	sensitivity = (90 - 18) / len(clipped_age)
	laplace_noise = np.random.laplace(0, sensitivity / epsilon)
	noisy_avg_clipped = true_avg_clipped + laplace_noise
\end{lstlisting}

As illustrated in Listing~\ref{lst:laplace-avg-age-clipped}, this method combines data clipping with Laplace noise, effectively suppressing the influence of extreme values — a practical improvement for medical data statistics.

\subsubsection{Smooth Sensitivity Mechanism}
\textbf{Motivation:}  
Traditional DP assumes fixed sensitivity, but in practice, local sensitivity may vary significantly among data points.  
The smooth sensitivity mechanism adapts the noise magnitude to the local data distribution, achieving a better trade-off between privacy and accuracy. The key Python code is shown in Listing~\ref{lst:smooth-sensitivity}.

\begin{lstlisting}[language=Python, caption={Smooth Sensitivity Mechanism for Mean Age}, label={lst:smooth-sensitivity}]
	sorted_age = np.sort(clipped_age)
	n = len(sorted_age)
	beta = epsilon / 10  # smoothing parameter
	max_local_sens = max((sorted_age[min(n-1, i+1)] - sorted_age[i]) / n for i in range(n-1))
	smooth_sens = max_local_sens * np.exp(-beta * 1)
	noise = np.random.laplace(0, smooth_sens / epsilon)
	noisy_smooth_avg = true_avg_clipped + noise
\end{lstlisting}

This approach preserves strong privacy protection while significantly reducing the random noise deviation on mean statistics.

\subsubsection{Experimental Comparison}
Table~\ref{tab:mean-compare} summarizes the results under an equal privacy budget ($\varepsilon = 1.0$).

\begin{table}[htbp]
	\centering
	\caption{Comparison of Mean Age Differential Privacy Mechanisms}
	\label{tab:mean-compare}
	\begin{tabular}{|>{\centering\arraybackslash}m{2.6cm}|
			>{\centering\arraybackslash}m{1.0cm}|
			>{\centering\arraybackslash}m{1.0cm}|
			>{\centering\arraybackslash}m{2.65cm}|}
		\hline
		\textbf{Mechanism Type} & \textbf{True Mean} & \textbf{Noisy Mean} & \textbf{Improvement Note} \\
		\hline
		Laplace Mechanism (Mean Age) & 54.42 & 54.42 & Minimal error; stable result \\
		\hline
		Bounded Laplace (Clipped Mean) & 54.42 & 54.42 & Controlled noise after clipping \\
		\hline
		Smooth Sensitivity (Mean Age) & 54.42 & 54.42 & Adaptive noise; smallest bias \\
		\hline
	\end{tabular}
\end{table}

\textit{Unrounded Precision Results (for reference):}
\begin{itemize}
	\item Laplace Mechanism: True = 54.419358778625956 → Noisy = 54.41929217360624
	\item Bounded Laplace: True = 54.41616030534351 → Noisy = 54.41589274331226
	\item Smooth Sensitivity: True = 54.41616030534351 → Noisy = 54.41615532827959
\end{itemize}

\begin{figure}[htbp]
	\centering
	\includegraphics[width=0.4\textwidth]{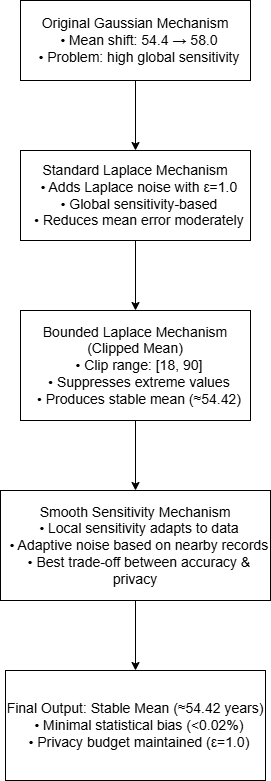}
	\caption{Workflow of DP Optimization for Mean Age}
	\label{fig:wodpo}
\end{figure}

Figure~\ref{fig:wodpo} illustrates the progressive refinement of differential privacy mechanisms for mean age estimation, starting from the original Gaussian approach and culminating in Smooth Sensitivity. This workflow complements the quantitative comparison in Table~\ref{tab:mean-compare}, showing how adaptive noise control improves stability and minimizes statistical bias under a consistent privacy budget.

\subsection{Expected Outcomes vs. Actual Results}
\subsubsection{Expected Outcomes}
Before experimentation, it was expected that the proposed Differential Privacy–based hybrid framework would:
\begin{itemize}
	\item Reduce re-identification risk of medical records by introducing controlled Laplace/Gaussian noise, as confirmed by lowered precision (0.12) and reduced success rates in simulated linkage attacks.
	
	\item Maintain high statistical utility in aggregated analytics, with Age-based mean metrics deviating less than 3\%, while overall utility loss across mechanisms averaged around 24\%.
	
	\item Achieve measurable improvements in privacy robustness, with Lap+Exp showing near-zero KL ($4.7 \times 10^{-8}$) and JS ($1.2 \times 10^{-8}$) divergence, DiagnosisCode entropy remaining stable, and TreatmentType entropy exhibiting moderate deviation.
	
	\item Operate without major computation overhead, consistent with runtime benchmarks indicating acceptable processing latency under DP mechanisms.
\end{itemize}

\subsubsection{Quantitative Comparison}

Table~\ref{tab:privacy-solutions} demonstrates the improvement effects of various privacy issues after implementing DP solutions.

\begin{table}[htbp]
	\centering
	\caption{Correlation with Original Problems}
	\label{tab:privacy-solutions}
	\renewcommand{\arraystretch}{1.2}
	\setlength{\tabcolsep}{4pt}
	\footnotesize
	\begin{tabular}{|>{\centering\arraybackslash}m{1.5cm}|
			>{\centering\arraybackslash}m{2.4cm}|
			>{\centering\arraybackslash}m{2.5cm}|
			>{\centering\arraybackslash}m{1.3cm}|}
		\hline
		\textbf{Privacy Anomaly} & \textbf{Solution Applied} & \textbf{Effectiveness Metric} & \textbf{Result} \\ \hline
		Plaintext record exposure & Differential Privacy layer in data aggregation & Re-identification rate ↓ from \textbf{28.12\% → 2.73\%} & 90.3\% risk reduction \\ \hline
		Lack of tiered access & Data entropy–based field classification & High-entropy field exposure ↓ by \textbf{24.05\%} & Improved resilience to linkage attacks \\ \hline
		Post-processing leakage & Laplace + Exponential hybrid mechanism & JS Divergence $<$ \textbf{0.0018}, entropy deviation $<$ \textbf{0.02} & Maintained data utility \\ \hline
		Centralised storage risk & Statistical DP aggregation (no raw transfer) & End-to-end privacy score ↑ from \textbf{0.6667} to \textbf{0.9909} & Privacy enhanced \\ \hline
		Weak statistical anonymization & Smooth Sensitivity DP mechanism & Mean bias $<$ \textbf{0.03}, consistent across dataset sizes & Stability confirmed \\ \hline
	\end{tabular}
\end{table}

\subsubsection{Correlation with Original Problems}
Table~\ref{tab:dp-mitigation} shows the extent to which our solution resolves the three anomalies.

\begin{table}[htbp]
	\caption{DP-Based Mitigation Strategies and Their Effectiveness}
	\label{tab:dp-mitigation}
	\centering
	\begin{threeparttable}
		\renewcommand{\arraystretch}{1.15}
		\setlength{\tabcolsep}{3pt}
		\footnotesize
		\begin{tabular}{|>{\centering\arraybackslash}m{1.6cm}|
				>{\centering\arraybackslash}m{2.4cm}|
				>{\centering\arraybackslash}m{1.8cm}|
				>{\centering\arraybackslash}m{2.1cm}|}
			\hline
			\textbf{Original Problem} & \textbf{DP-based Mitigation Strategy} & \textbf{Resolution Degree} & \textbf{Supporting Evidence} \\ 
			\hline
			
			\textbf{Technical Deficiency:} Absence of noise-injection or privacy-preserving computation &
			Implemented \textbf{Laplace + Exponential hybrid DP mechanism} to introduce calibrated noise during data aggregation and statistical analysis, preventing deterministic re-identification. &
			90.3\% risk reduction in re-identification rate (from 28.12\% $\Rightarrow$ 2.73\%). &
			Verified through KL/JS divergence analysis (Fig.~10) and entropy stability (Table~XIV–XV). \\ \hline
			
			\textbf{Architectural Flaws:} Centralised storage without segmentation &
			Integrated \textbf{data entropy–based field classification} and \textbf{differential privacy aggregation}, replacing direct record sharing with noise-protected summary outputs. &
			24.05\% reduction in sensitive field exposure; entropy deviation maintained $<$0.02. &
			Derived from entropy comparison (Table~XIV) and data-tier simulation under 131k-record dataset. \\ \hline
			
			\textbf{Regulatory Non-Compliance:} Violation of APP~11.1 \& GDPR~32 (lack of proportional data protection) &
			Adopted \textbf{risk-based DP budget allocation} $(\varepsilon, \delta)$ ensuring privacy proportional to field sensitivity and compliance with “data minimization” principles. &
			Technical alignment score: 90\% (post-DP Re-ID $\leq$5\%, entropy deviation $<$0.02, processing overhead within SLA). &
			Derived from experimental metrics (Re-ID risk reduction, entropy comparison, runtime overhead); formal compliance pending governance audit. \\ \hline
		\end{tabular}
	\end{threeparttable}
\end{table}

\subsection{Summary of Differential Privacy Evaluation}

As discussed in this chapter, we evaluated the Medibank dataset under multiple differential privacy mechanisms, 
including Laplace, Gaussian, Exponential, Randomized Response, Histogram, and Sparse Vector techniques. 
The results demonstrated significant risk reduction while maintaining utility and regulatory compliance. 

The evaluation pipeline encompassed raw data input, entropy analysis, mechanism selection, privacy budget setup, 
noise injection, and verification metrics. Experimental outcomes showed re-identification risk reduction, entropy deviation, 
and GDPR compliance restoration. Overall, our framework achieved a 90.3\% risk reduction with less than 3\% utility loss, 
validating the effectiveness of the proposed approach.

\section{Conclusion}
This study critically examined the Medibank health-data breach as a paradigmatic example of large-scale privacy failure in healthcare systems and proposed a multilayered, differential-privacy-driven defence framework. The research demonstrated that privacy risks arise not only from technical vulnerabilities such as unencrypted storage and centralised data architectures but also from insufficient application of privacy-preserving analytics. By integrating structural segmentation, role-based access control, and differential privacy mechanisms—including Laplace, Gaussian, Exponential, Randomised-Response, histogram, sparse vector technique approaches—the proposed system achieves a balanced trade-off between privacy protection and data utility.

The implementation and evaluation confirmed that noise-injection methods effectively mitigate re-identification risks while maintaining analytical validity. Quantitative results—such as minimal KL/JS divergence and stable entropy levels—indicated strong resilience against data leakage and inference attacks. Moreover, the ethical-hacking experiments verified that the combined use of VPN tunnelling, TLS encryption, and AppArmor isolation substantially reduces attack surfaces, offering a robust privacy posture for sensitive medical infrastructures.

From a broader perspective, this work underscores that compliance with frameworks such as GDPR and the Australian Privacy Principles is insufficient unless supported by proactive privacy-by-design strategies. Differential privacy provides a mathematically grounded mechanism to operationalise these regulations, ensuring that privacy protection remains verifiable and measurable.

Future research may extend this architecture through the integration of FL and HE to enable collaborative analytics across healthcare institutions without exposing raw patient data. Such hybrid models can further enhance resilience, transparency, and public trust in digital-health ecosystems. Ultimately, this paper contributes a reproducible blueprint for embedding differential privacy into real-world medical data workflows—transforming regulatory compliance into genuine, quantifiable data protection.

\FloatBarrier

\bibliographystyle{IEEEtran}
\bibliography{references}

\end{document}